\documentclass[aps,prb,twocolumn,superscriptaddress,floatfix]{revtex4-1}
\usepackage{graphicx}
\usepackage{dcolumn}
\usepackage{bm}
\usepackage{stmaryrd}
\usepackage{latexsym}
\usepackage{amssymb}
\usepackage{amsfonts}
\usepackage{amsmath}
\usepackage{fancybox}
\usepackage{color}
\usepackage{verbatim}
\usepackage{bbold}

\def\be{\begin{equation}}
\def\ee{\end{equation}}
\def\ber{\begin{eqnarray}}
\def\eer{\end{eqnarray}}

\begin{document}
%

\title{ Flatbands in twisted double bilayer graphene}

\author{Narasimha Raju Chebrolu}
\affiliation{Department of Physics, University of Seoul, Seoul 02504, Korea}

\author{Bheema Lingam Chittari}
\affiliation{Department of Physics, University of Seoul, Seoul 02504, Korea}

\author{Jeil Jung}
\email{jeiljung@uos.ac.kr}
\affiliation{Department of Physics, University of Seoul, Seoul 02504, Korea}

\begin{abstract}
Flatbands with extremely narrow bandwidths on the order of a few mili-electron volts can appear in twisted multilayer graphene systems for appropriate system parameters. 
Here we investigate the electronic structure of a twisted bi-bilayer graphene, or twisted double bilayer graphene, to find the parameter space where isolated flatbands can emerge as a function of twist angle, vertical pressure, and interlayer potential differences. 
We find that in twisted bi-bilayer graphene the bandwidth is generally flatter than in twisted bilayer graphene by roughly up to a factor of two in the same parameter space of twist angle $\theta$ and interlayer coupling $\omega$, making it in principle simpler to tailor narrow bandwidth flatbands. Application of vertical pressure can enhance the first magic angle in minimal models at $\theta \sim 1.05^{\circ}$ to larger values of up to $\theta \sim 1.5^{\circ}$ when $ P \sim 2.5$~GPa, where $\theta \propto \omega/ \upsilon_{F}$.
Narrow bandwidths are expected in bi-bilayers for a continuous range of small twist angles, i.e. without magic angles, when intrinsic bilayer gaps open by electric fields, or due to remote hopping terms.
We find that moderate vertical electric fields can contribute in lifting the degeneracy of the low energy flatbands by enhancing the primary gap near the Dirac point and the secondary gap with the higher energy bands.
Distinct valley Chern bands are expected near $0^{\circ}$ or $180^{\circ}$ alignments.
\end{abstract}
\pacs{73.22.Pr, 71.20.Gj,31.15.aq}
\maketitle
%

\section{Introduction}
Research on twisted hybrid van der Waals 2D materials has seen recently a new surge of interest following 
experimental observations of exotic quantum phases due to strong electron correlations~\cite{Cao,kyounghwan} 
and especially signatures of unconventional superconductivity~\cite{pjarillo_superconductivity, Yankowitz,planckian} 
in twisted bilayer graphene (tBG),
raising hopes of finding new clues for understanding analogous behaviors seen in more complex systems.~\cite{complexoxides}
In tBG the spatial variation of interlayer coupling modifies the intrinsic Dirac cone band structure of graphene in such a way
that the band dispersion is almost completely suppressed at the so called magic twist angles.~\cite{bistritzer2011}
When the bandwidth $W$ of these low energy bands is sufficiently narrow it is possible to achieve the 
$U/W \gtrsim 1$ condition that makes the effective Coulomb repulsion $U$ more dominant.
A considerable body of literature has formed recently on the Coulomb interaction driven broken symmetry 
phases~\cite{Ochi, Pizarro, Venderbos, Phillips2018,koshino_wannier, Kang1, Kang2} 
and superconductivity in tBG flatbands~\cite{Fidrysiak, Roy, Huang, Tanmoy, Liu, Peltonen, Kennes, Isobe, Vishwanath, Vishwanath2, Wu, Guinea, Gonzalez, Su, Lian, Laksono, Tang, Dante1} 
in an effort to elucidate the nature of the superconducting phases.
Analogous observations of Coulomb interaction driven correlated phases and superconductivity have been observed
in ABC trilayer graphene (TG) on hexagonal boron nitride (hBN)~\cite{guorui_mott2018,guorui_superconductivity2019,guorui2019b} 
where the flattening of the low energy bands is facilitated by the presence of a vertical electric field that introduces a band gap
at the primary Dirac point of a chiral two-dimensional electron gas (2DEG),~\cite{prljung,prlzhang}
or in twisted gapped Dirac materials.~\cite{Dante2,javvaji,mos2bilayers}
It was suggested that the low energy flatbands~\cite{footnote} 
could have well defined valley Chern numbers and give rise to spontaneous quantum Hall phases when the band degeneracy is lifted by Coulomb interactions.~\cite{todadri,prljung}
The proposals of flatbands in several types of multilayer graphene materials
is suggesting that they can arise in a large variety of 2D material combinations 
provided that we choose the appropriate intrinsic electronic structure of each layer and their interlayer coupling.~\cite{todadri}
In this work we study the flatband bandwidth phase diagram of twisted BG/BG system, that we refer to as
twisted bi-bilayer graphene (tBBG) or twisted double bilayer graphene, which consists of two bilayer graphene units with a twist.
We assess for this system the effect of the interlayer coupling strength and the interlayer potential differences 
between the layers in the resulting bandwidth of the low energy flatbands.
It is expected that the smaller parabolic band dispersion slopes at low energy in a BG can favor the 
formation of flatbands upon interlayer hybridization.
This manuscript is structured as follows. In Sec.~II we introduce the theoretical details of 
the continuum model Hamiltonian used to formulate the problem. 
In Sec.~III we present the phase diagram of $U/W$, the flatband bandwidth and gaps as a function of different system
parameters, such as the twist angle, the interlayer coupling strength, and the interlayer potential differences due to a vertical electric field. 
In Sec.~IV we discuss the valley Chern number phase diagrams, and then close the paper in Sec.~V with the summary and conclusions.

\begin{figure}
\begin{center}
\includegraphics[width=8.5cm]{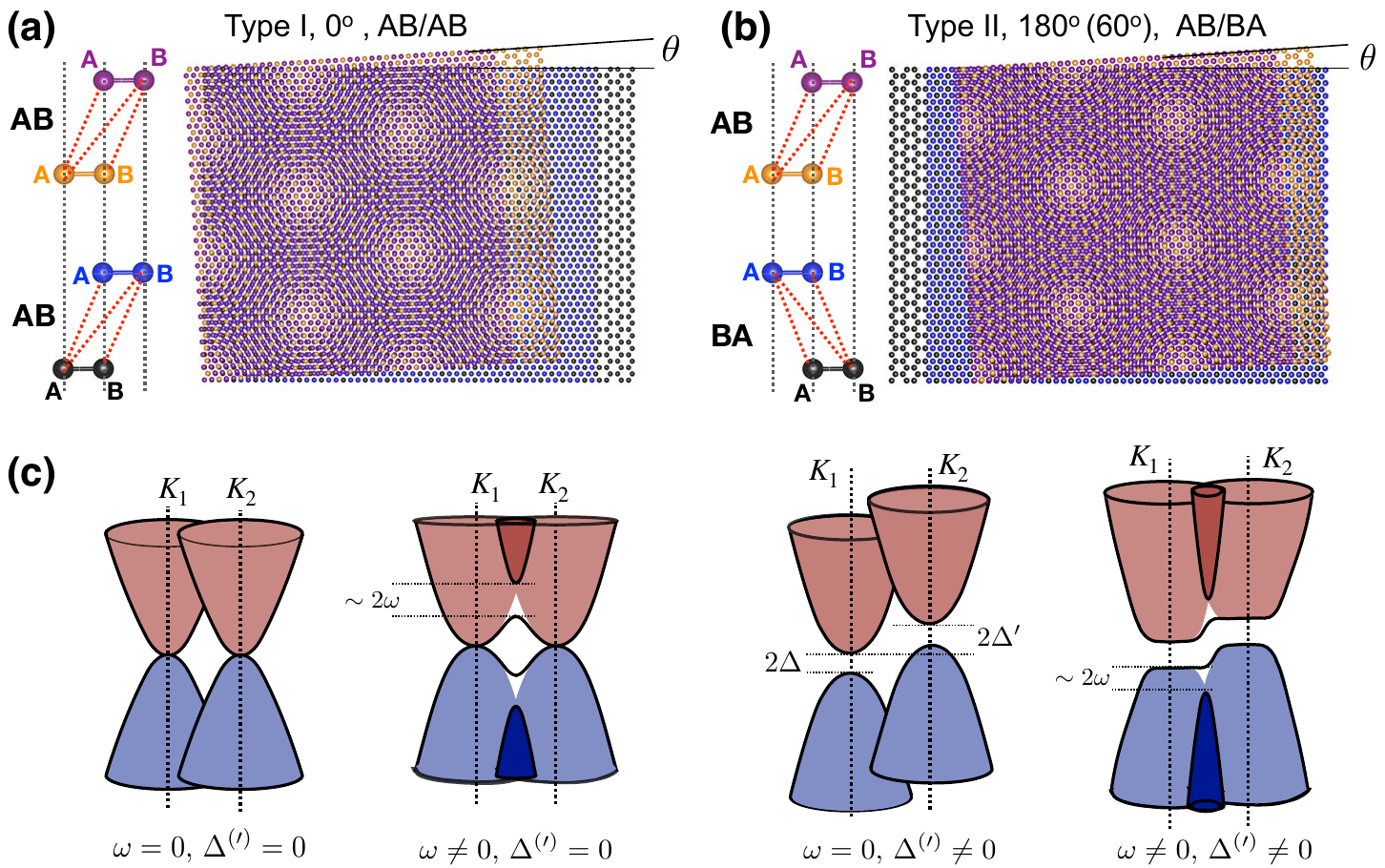}
\end{center}
\caption{
(Color online). Moire pattern created by a twisted bi-bilayer graphene (tBBG) and the 
commensurate unit cell of two Bernal stacked bilayer graphene aligned (a) near $0^{\circ}$ twist angle for type I AB/AB twisted bi-bilayers, 
and (b) near 180$^{\circ}$  (or 60$^{\circ}$) twist angles for for type II AB/BA twisted bi-bilayers.
The panel (c) represents schematically two uncoupled parabolic bands, the perturbative interband hybridization through the 
interlayer tunneling $\omega$, the intrinsic bilayer gap ($2\Delta$), and band offset ($2\Delta^{\prime}$) 
due to a perpendicular electric field.
}
\label{figure1}
\end{figure}
\begin{figure*}
\begin{center}
\includegraphics[width=17cm]{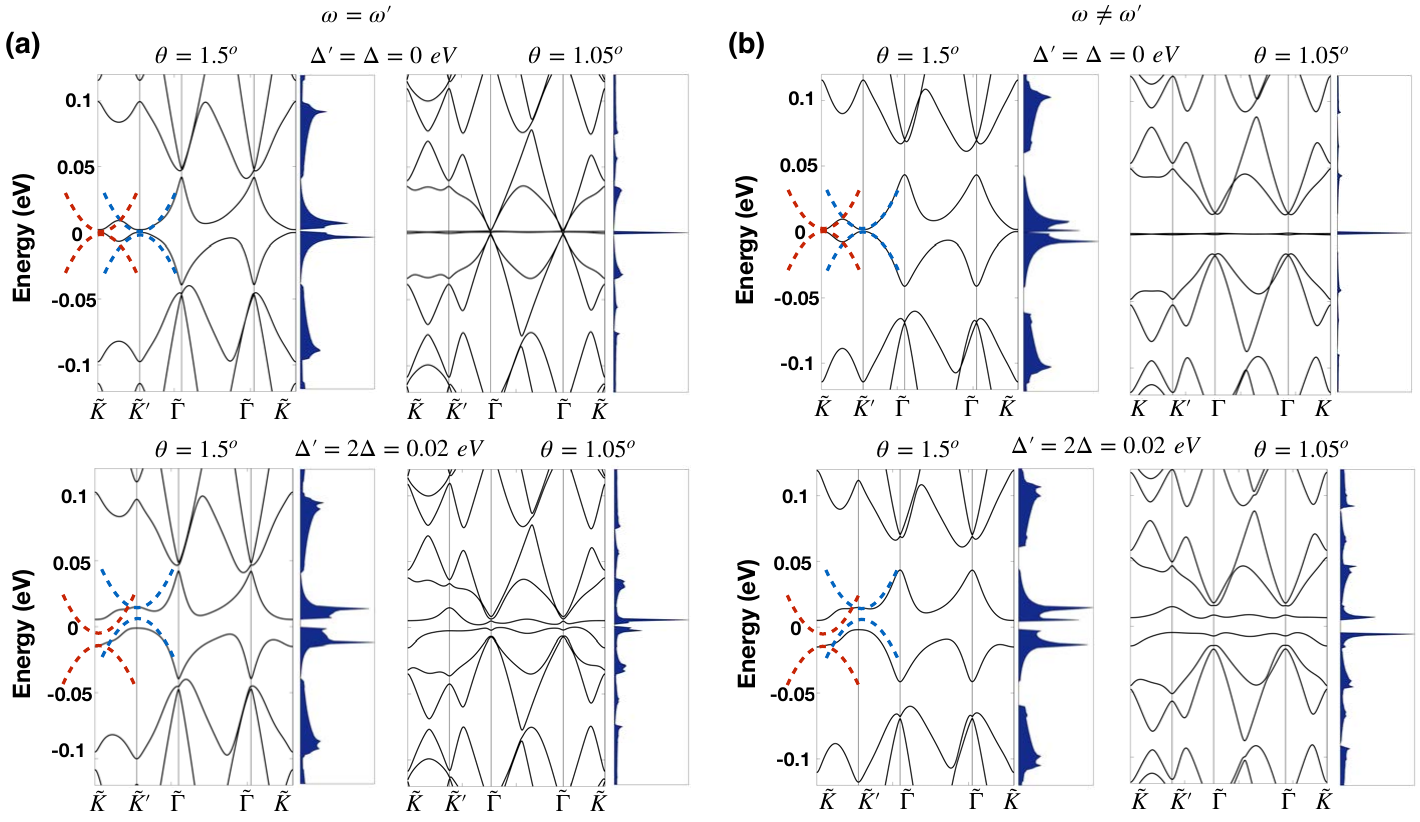}
\end{center}
\caption{
Band structures of bi-bilayer graphene for $\theta=1.5^{\circ}, 1.05^{\circ}$, 
for zero and finite interlayer potential difference parameters $\Delta$, $\Delta'$. 
The parabolic dotted lines are a guide to the eye for the original position of the bilayer graphene band edge
in the absence of interlayer tunneling between the BG. 
The addition of an interlayer potential difference through the $\Delta^{\prime}$ parameters 
introduces a separation between the low energy flatbands roughly proportional to the interlayer potential
difference between the top and bottom outer layers of tBBG. 
For a system with Fermi velocity $\upsilon_{F} = 1.0\times 10^{6}$~m/s we show the 
(a) band structures calculated with a single parameter interlayer coupling $\omega=\omega'=0.12$~eV,
and (b) band structures calculated with $\omega'=0.1$ and $\omega=0.12$~eV with different inter-sublattice tunneling values
that enhances the gap between the flatbands and the neighboring higher energy bands.
 }
\label{figure2}
\end{figure*}
%

%
%
\begin{figure}
\begin{center}
\includegraphics[width=8.5cm]{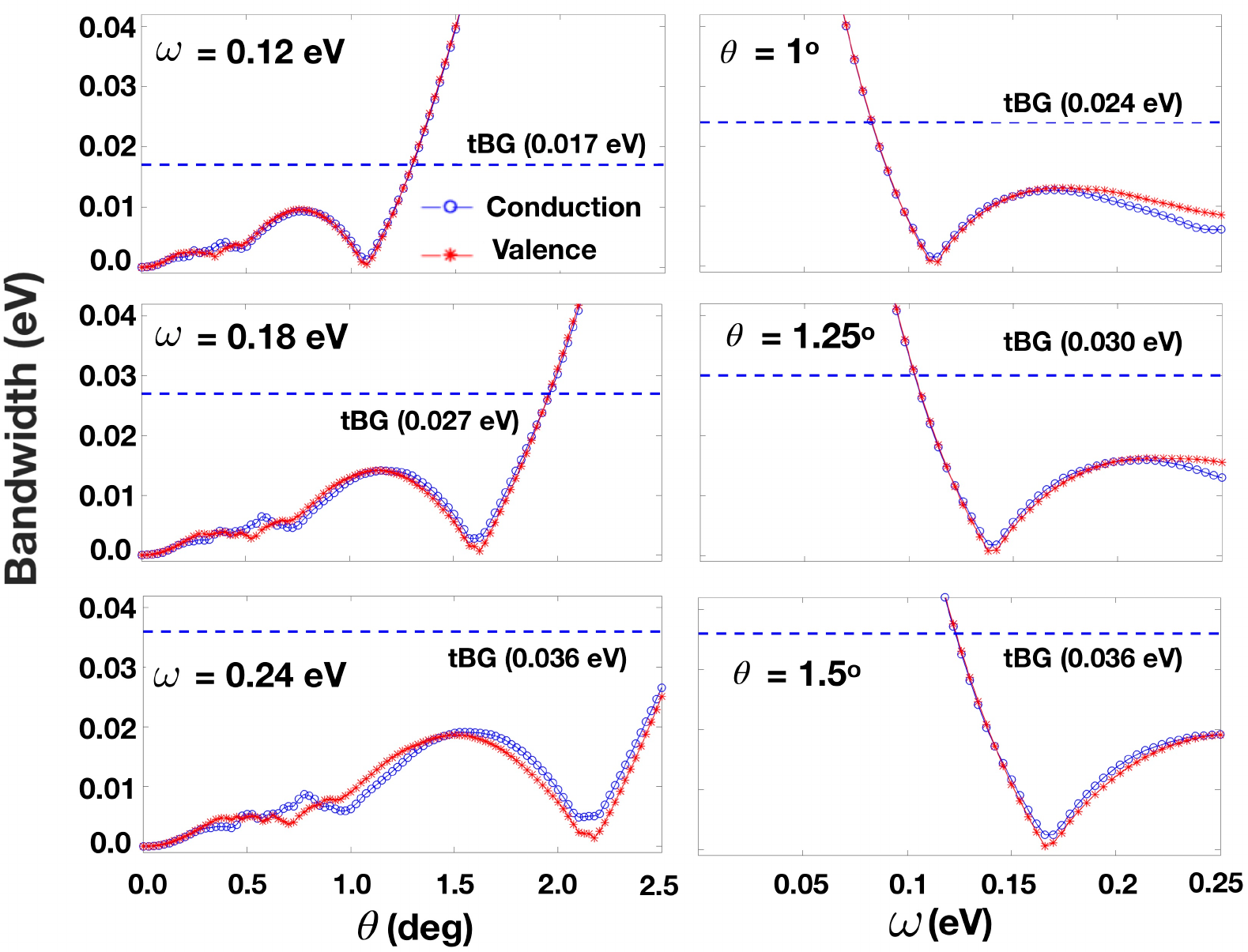}
\end{center}
\caption{
(Color online)  Variation of flatband bandwidth for $\Delta=\Delta^{\prime}=0 $~eV as a function of inter-layer 
coupling strength and twist angle for the rigid continuum model with $t_0 = -2.6$~eV.
The bandwidths in tBBG are narrower by roughly a factor of two when compared with the bandwidths in tBG with similar system parameters,
where the dotted horizontal lines represent the bandwidth of tBG at the belly maxima.
{\em Left Panel:} Flatband bandwidth as a function of twist angle $\theta$ at different inter-layer couplings 
$\omega = 0.12, 0.18, 0.24$~eV. We observe that the bandwidth changes non-monotonically with the twist angle and 
goes through a series bandwidth minima. 
{\em Right Panel:} Flatband bandwidth as a function of $\omega$, at different twist 
angles $\theta^{\circ}=1^{\circ}, 1.25^{\circ}, 1.5^{\circ}$.
We can observe a steep initial reduction in the bandwidth followed by a mild bump for increasing $\omega$.
 }
\label{figure3}
\end{figure}
\begin{figure*}
\begin{center}
\includegraphics[width=17.5cm]{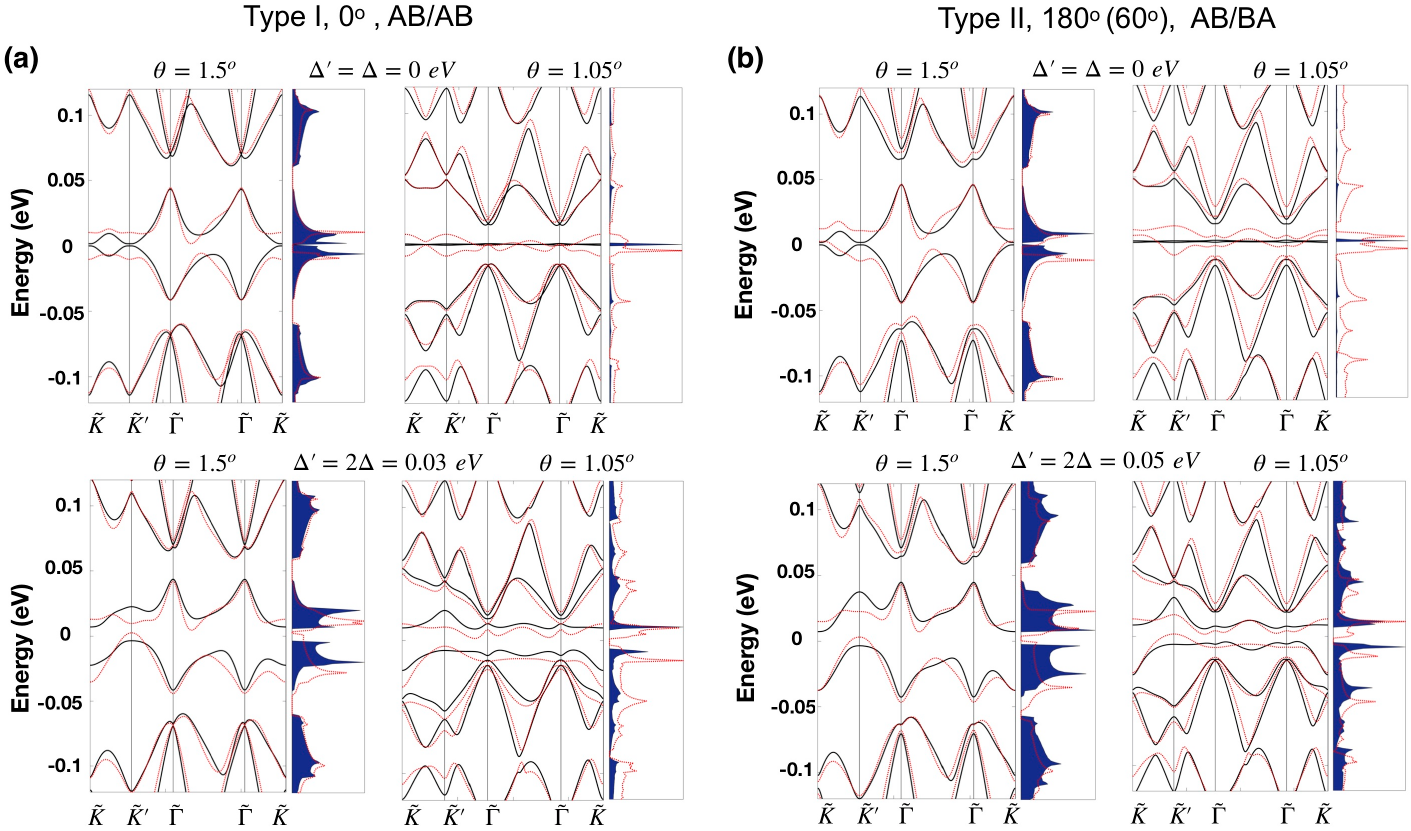}
\end{center}
\caption{
Comparison of the minimal and remote hopping Hamiltonian model that includes the  
trigonal warping $t_3$, intrinsic electron-hole asymmetric $t_4$ terms and higher energy dimer site potential $\delta$.
The band-structures of the minimal model are in black, and those that include the remote hopping terms are in dotted red lines.
The remote hopping terms generally widen the minimal model flatbands, and can 
introduce primary band gaps $\delta_p$ near charge neutrality for large twist angles.
Appropriate electric fields can compress the bandwidth, while maintaining band isolation through the primary $\delta_p$ and secondary gaps $\delta_s$.
We distinguish the (a) band structures for type I near 0$^{\circ}$ alignment and (b) for type II near 180$^{\circ}$ alignment 
that show distinct electronic structures and responses to electric fields.
}
\label{figure4remote}
\end{figure*}

\begin{figure*}
\begin{center}
\includegraphics[width=17cm]{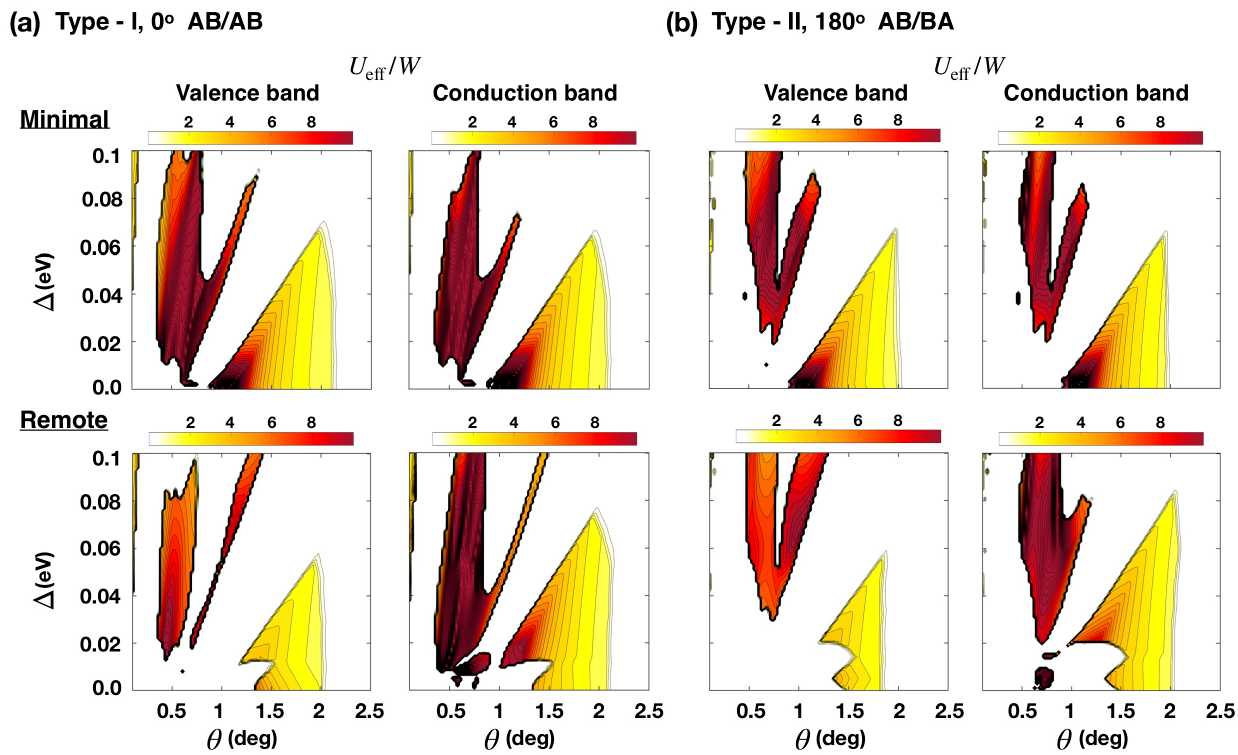}
\end{center}
\caption{
(Color online). Colormap phase diagram of $U_{\rm eff}/W$ for the effective Coulomb interaction versus bandwidth $W$
 in the parameter space of $\theta$ and  $\Delta$ that indicates the plausible regions where Coulomb interactions can trigger ordered phases.
 Inclusion of remote hopping terms in the band Hamiltonian of bilayer graphene introduces particle-hole symmetry breaking
 generally favoring $U_{\rm eff}/W$ for the conduction bands  over the valence bands. 
 Islands in the phase diagram are found due to the suppression of $U_{\rm eff}$ in Eq.~(\ref{ueffeq}) when neighboring bands overlap,
 leading to large regions of twist angles between $0.5^{\circ} - 0.8^{\circ}$ and $1^{\circ} - 1.6^{\circ}$
 favorable for interaction driven broken symmetry phases for $\Delta$ values accessible in experiments.
}
\label{ratio}
\end{figure*}

\begin{figure*}
\begin{center}
\includegraphics[width=17.5cm]{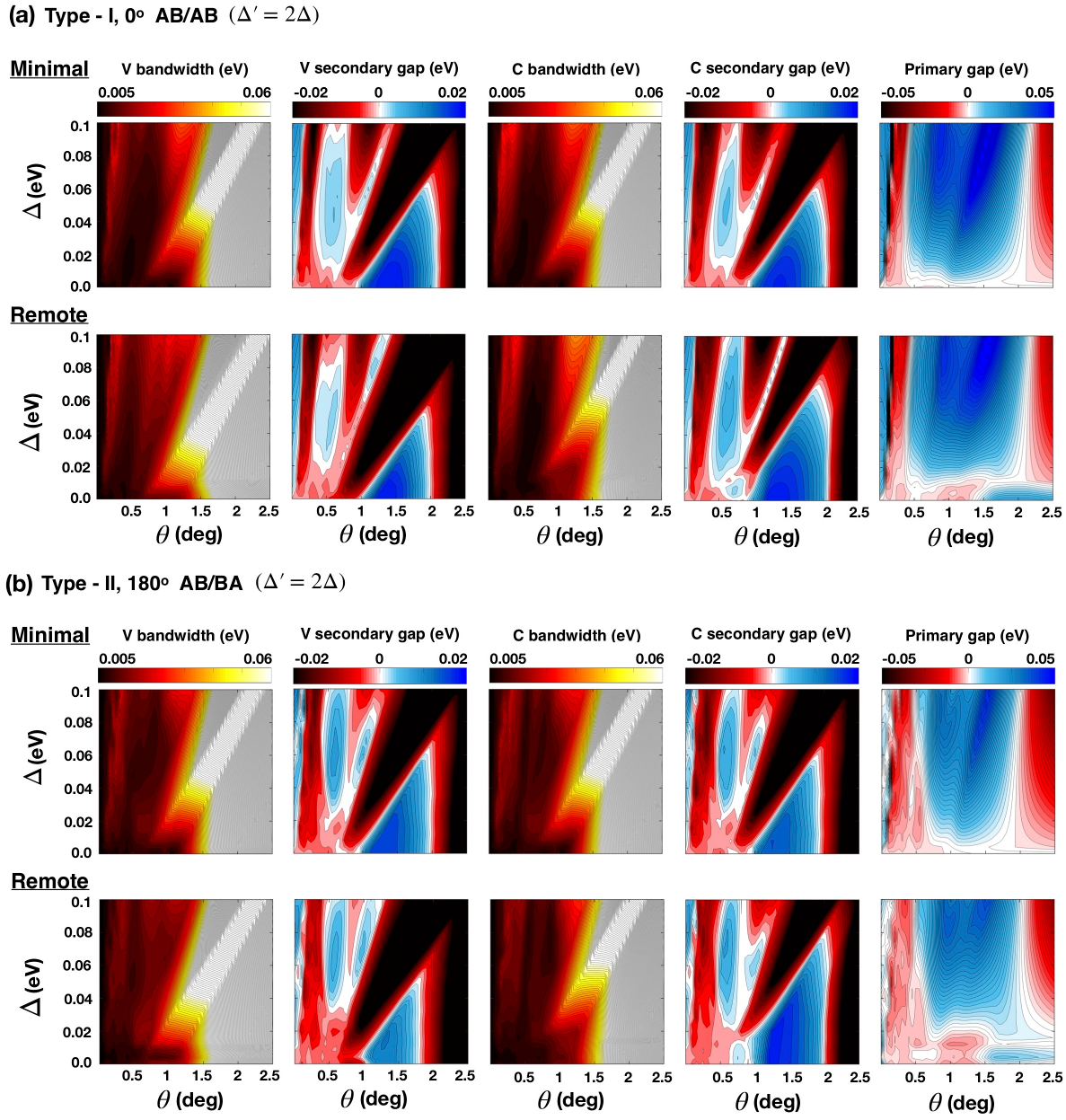}
\end{center}
\caption{
(Color online)  Variation of flatband bandwidth and band isolation through the primary $\delta_p$ and secondary $\delta_s$ band gaps
for the low energy valence (V) and conduction (C) bands near charge neutrality as a function of twist angle $\theta$ and interlayer potential difference $\Delta$,
shown for (a) type-I near $0^{\circ}$ alignment as a function of $\theta$ and $\Delta$, and for (b) type-II near $180^{\circ}$ alignment, that requires a larger $\Delta$ to achieve positive primary gaps $\delta_p$ for $\theta \gtrsim 1^{\circ}$ to simultaneously achieve $\delta_{p/s} >0$.
The remote hopping terms introduce particle-hole symmetry breaking that expands the parameter space where the low energy flatbands 
are isolated and generally favors the isolation of the conduction band over the valence bands. Negative values for the gaps indicate overlap with neighboring bands. 
In the presence of remote hopping terms, simultaneous $\delta_{p/s}>0$ are expected in the conduction bands 
for small $\Delta$ in the vicinity of $\theta \sim 0.7^{\circ}$  and in particular
around $\theta \gtrsim 1.5^{\circ}$, for both near $0^{\circ}$ and $180^{\circ}$ alignments.
These regions should be more accessible with scanning probe measurements 
where vertical electric fields and induced carrier densities are less directly controllable than in top-bottom dual gated devices.
 }
\label{figure5n}
\end{figure*}

\begin{figure*}
\begin{center}
\includegraphics[width=17.5cm]{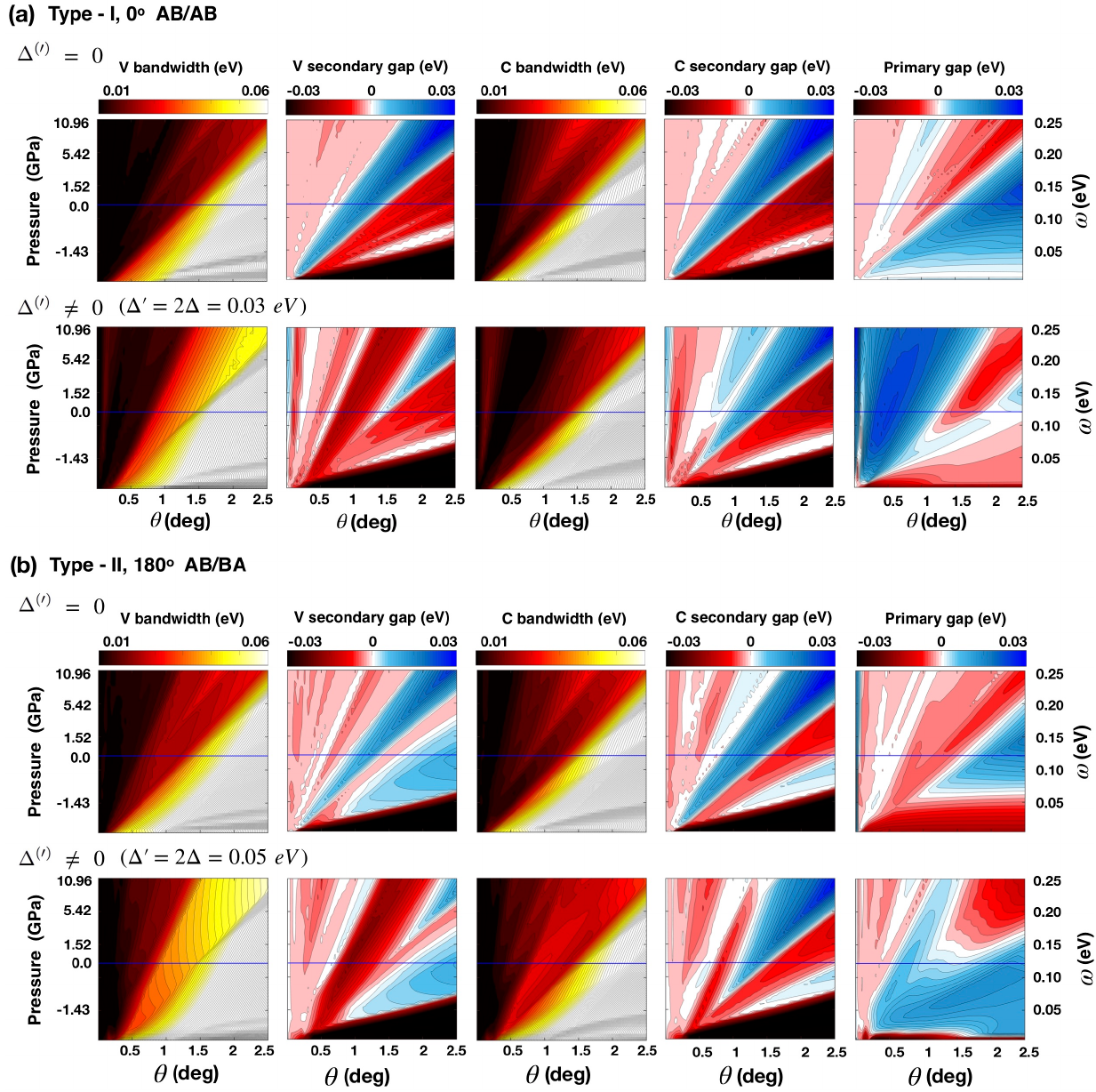}
\end{center}
\caption{
(Color online)
Phase diagram of bandwidth and band isolation of the flatband through primary $\delta_p$ and secondary $\delta_s$ gaps 
with the surrounding bands, calculated both for the valence and conduction flatbands
as a function of $\theta$ and $\Delta^{\prime} = 2 \Delta$
for (a) type-I near $0^{\circ}$ alignment , and for (b) type-II near $180^{\circ}$ alignments.
The Hamiltonian model for the above results includes the remote hopping terms and the calculations were carried out 
for $\Delta^{\prime} = 0$, and for finite interlayer potential $\Delta^{\prime} \neq 0$. 
More phase diagrams for other $\Delta$ values and calculations for the minimal model can be found in the supplemental material.~\cite{supplemental}
}
\label{figure_pressure}
\end{figure*}

\section{Model Hamiltonian for twisted bi-bilayer graphene}    
Models proposed in the literature to capture the electronic structure of tBG 
relied either on tight-binding calculations~\cite{lopes,shallcross,koshino} 
often based on the distance dependent two center approximation models for the hopping terms between interlayer carbon atoms, 
or by using other more sophisticated parametrizations.~\cite{jung2014,kaxirasparam} 
The successful formulation of a rigorous moire bands theory on the basis of the moire pattern superlattice~\cite{bistritzer2011} 
allows to obtain accurate continuum models for the Hamiltonian informed from first principles calculations.~\cite{jung2014}
In the present work we extend the continuum model of
Bistritzer-MacDonald model for the tBG~\cite{bistritzer2011} to the case of tBBG.
The Hamiltonian of tBBG at valley $K$ that we use captures the interlayer coupling between
the twisted layers through a first harmonic stacking-dependent interlayer tunneling function. 
Schematic representations of commensurate and twisted BG/BG structures are shown in 
Fig.~\ref{figure1}.
We write the Hamiltonian of twisted top ($+$) and bottom ($-$) bilayer graphene Hamiltonian subject to $\Delta_i$ intralayer potentials as
\begin{equation}
H_{\rm tBBG} (\theta) = 
\begin{pmatrix} 
h^+_{t}  + \bar{\Delta}_1  & t^{+}_{s}  &  0   & 0  \\ 
t^{+ \dag}_{s}  & h^+_{b} + \bar{\Delta}_2   &  T({\bf r})   & 0  \\ 
0  &  T^{\dag}({\bf r})  &  h^-_{t}  + \bar{\Delta}_3    &   t^{-}_{ s}  \\
0  & 0 &  t^{- \dag}_{ s}   &   h^-_{b} + \bar{\Delta}_4  \\ 
\end{pmatrix},
\label{hamiltonian}
\end{equation}
where $h^{\pm}_{t/b} = h_{t/b}(\pm \theta / 2)$ such that the relative twist angle between the bilayers is $\theta$.
The top and bottom BG are labeled through the positive/negative ($+/-$) rotation signs,
while in turn we have top/bottom ($t/b$) graphene layers within each BG that are coupled through the matrices $t_s^{\pm}$ that we define later.
The site potentials for each graphene layer $\Delta_i$ are mapped on its sublattices through 
$\bar{\Delta}_i = \Delta_i \mathbb{1}$ where $i=1,\,2,\,3,\,4$ are the layer labels from top to bottom, and $\mathbb{1}$ is a $2\times2$ identity matrix.
Potential differences can give rise to band gaps at the primary Dirac point of each BG and shift the associated band edges.
We will discuss later on the effects of these intralayer potentials in the electronic structure of the flatbands. 
The Hamiltonian of graphene is given by $h^{\pm}_{l}(\theta)= h^{\pm}(\theta) +  \delta \left( \mathbb{1} - l s \sigma_z \right)/2 $ where the second 
term adds a $\delta = 0.015$~eV sublattice potential at the higher energy dimer sites at the $t/b$ layers $l = \pm$,~\cite{accuratebilayer} 
that depends on AB or BA stacking $s = \pm$ respectively.
The Dirac Hamiltonian given by $h(\theta)= \upsilon_{F} \hat{R}_{-\theta}{\bf p} \cdot \sigma_{xy}$ includes a phase shift due to a rotation 
$\hat{R}_{-\theta}$ such that $e^{ \pm i \theta_{\bf p}} \rightarrow e^{\pm i (\theta_{\bf p} - \theta)} $, 
where $\sigma_{xy} = ( \sigma_x, \sigma_y)$ and $\sigma_z$ are the graphene sublattice pseudospin Pauli matrices, 
and the momentum is defined in the $xy$-plane ${\bf p} = (p_x, p_y)$, 
where we assume $K$ valley unless stated otherwise. 
The Fermi velocity $\upsilon_{\rm F} = \upsilon_0$ defined from $ \upsilon_i = \sqrt{3} \left| t_i \right| a/2 \hbar $
is related to the intralayer nearest neighbor hopping term $t_0= - 2.6$~eV within the local density approximation (LDA),~\cite{ldahopping}
while an enhanced $t_0 = -3.1$~eV and ab initio interlayer tunning captures better the experimental moire band features.~\cite{dillon2015}
The interlayer coupling model of a bilayer graphene is given by 
\begin{eqnarray}
t^{\pm}_{\rm AB} =  
 \begin{pmatrix}   - \upsilon_4 \pi ^{\pm \dag}   & - \upsilon_3 \pi^{\pm}  \\   t_1   & -\upsilon_{4} \pi^{\pm \dag} \end{pmatrix}, \quad
t^{\pm}_{\rm BA} =  t^{\pm \dag}_{\rm AB}
\end{eqnarray}
satisfying $t_{s=+} = t_{s=-}^{\dag}$ for 
AB or BA ($ s = \pm 1$) stacking dependent interlayer coupling 
that consists of a minimal coupling term $t_{ s} = t_1 (\sigma_x  -  i s \sigma_y)/2$ plus 
remote hopping contributions through $t_3= 0.283$, $t_4= 0.138$ terms, 
giving rise to trigonal warping and electron-hole asymmetry.
The $\pi^{\pm}$ operators include the phases due to $\pm \theta/2$ layer rotation.
The type II AB/BA bi-bilayers near 180$^{\circ}$ alignment can be modeled by controlling the stacking of bottom BG from AB to BA by using $s=-1$ for the bottom BG. 
The interlayer tunneling is defined as the Hamiltonian matrix element at the Dirac point $t_1 = H_{BA^{\prime}}(K, \vec{d}_{\rm AB}) $ 
between $B$ to $A^{\prime}$ sites from bottom to top layer for AB stacking when both atomic sites are vertically aligned
and assume $t_1=0.361$~eV at zero pressure within LDA.~\cite{accuratebilayer}
We can identify the interlayer tunneling with the first harmonic expansion coefficient of the interlayer coupling such that 
$t_1 = 3 \omega$,~\cite{jung2014} and for simplicity we use the same AB stacking tunneling within each Bernal BG and the twisted interfaces.
The minimal model approximation uses $\delta= t_3 = t_4 = 0$ in Eq.~\ref{hamiltonian}.
The presence of remote hopping terms will lead to broadening of the low energy flatbands and enhancemenet in electron-hole asymmetry.
This behavior is not strange since the $t_3$ trigonal warping widens the range of band touching 
points at three points away from the Dirac point at directions connecting the $K$ points with $\Gamma$,~\cite{accuratebilayer}
and the $t_4$ term breaks the intrinsic electron hole symmetry of bilayer graphene.~\cite{mccann}

The moire Brillouin zone (mBZ) orientation is preserved when the top and bottom graphene layers rotate symmetrically in opposite senses.
In the small angle approximation the interlayer coupling Hamiltonian is given by
\begin{eqnarray}
T({\bf r}) = \sum_{j=0,\pm} e^{ -i {\bf Q}_j {\bf r}} T^{j}_{l, l'},   \label{interlayercoupling}
\end{eqnarray}
where the three ${\bf Q}_j$ vectors ${\bf Q}_0 = K_D \theta (0, -1) $ and 
${\bf Q}_{\pm} = K \theta (\pm \sqrt{3}/2, 1/2)$ are proportional to twist angle $\theta$ and 
$K_D = 4 \pi / 3 a$ is the Brillouin zone corner length of graphene, whose lattice constant is $a=2.46~\AA$,
and here the indices $l, \, l'$ label the sublattices of neighboring twisted surface layers.
The interlayer coupling matrices between the two rotated adjacent layers are given by
\begin{equation}
T^0 =   \begin{pmatrix} \omega^{\prime}  & \omega  \\ \omega  &  \omega^{\prime} \end{pmatrix},   \,\, \,\,   
T^{\pm} =  \begin{pmatrix} \omega^{\prime}  & \omega e^{\mp i 2\pi/3}  \\ \omega e^{\pm i 2\pi/3}  &  \omega^{\prime}  \end{pmatrix}
\end{equation}
using a form that distinguishes interlayer tunneling matrix elements $\omega = \omega_{BA^{\prime}}$ and $\omega^{\prime} = \omega_{AA^{\prime}}$ 
for different and same sublattice sites between the layers.
The convention taken here for the $T^{j}$ matrices~\cite{jung2014}  assume an initial AA stacking configuration $\tau = (0,0)$
and differs by a phase factor with respect to the initial AB stacking $\tau = (0,a/\sqrt{3})$.~\cite{bistritzer2011}
The greater interlayer separation $c $  compared to the carbon-carbon distances $ a_{\rm CC}$ 
lead to slowly varying interlayer tunneling function $T({\bf r})$ and the moire patterns can often be accurately described within a first harmonic approximation~\cite{bistritzer2011,jung2014}.
In this limit, and assuming no corrugation effects, the interlayer coupling strength can be well approximated 
by a single parameter $\omega = \omega^{\prime}$ whose value was calculated within the local density approximation (LDA) 
to be $\omega \sim 0.113$~eV when averaged for every stacking at a fixed interlayer distance $c_{\rm AB}=3.35~\AA$  of AB stacking.
A somewhat weaker $\omega \sim 0.098$~eV is expected when the interlayer relaxations for farther AA interlayer distance $c_{\rm AA} = 3.57~\AA$ within LDA
is accounted for in the averaging process.~\cite{jung2014}
The interlayer tunneling matrix elements $H_{ll^{\prime}}(K, \vec{d})$ are evaluated at the Dirac point $K$ 
for a commensurate system with stacking sliding vector $\vec{d}$ through 
the lattice Fourier transform of the distant real space hopping terms connecting the sites $l$ and $l'$.
The tunneling matrix elements $\omega$, $\omega^{\prime}$ for twisted systems are obtained averaging over all
possible commensurate stacking configurations given by the integral
\begin{eqnarray}
\omega_{ll^{\prime}} = \int_{\rm cell} {\rm d} \vec{d} \, H_{ll^{\prime}}(K, \vec{d}) 
\simeq \sum_{s} \frac{ H_{ll^{\prime}} (K, \vec{d}_{s})}{3} 
\end{eqnarray}
that in the first harmonic approximation can be approximated by taking the average of the sum over the
three symmetric stacking configurations $s = {\rm AA, \, AB, \, BA}$ at their respective equilibrium interlayer distances.~\cite{jung2014}
Because tunneling between interlayer sublattices in graphene on graphene vanish at symmetric stackings $s$ when they are not vertically aligned we have
\begin{eqnarray}
\omega &=& \omega_{AB^{\prime}} = \omega_{BA^{\prime}} \simeq \frac{H_{BA^{\prime}}(K,d_{\rm AB})}{3} \label{omega} \\ 
\omega^{\prime} &=& \omega_{AA^{\prime}} = \omega_{BB^{\prime}} \simeq \frac{H_{AA^{\prime}}(K,d_{\rm AA})}{3}.  \label{omegap}
\end{eqnarray}
Using Eqs.~(\ref{omega}-\ref{omegap}) at zero pressure and using the EXX+RPA equilibrium distances for each stacking~\cite{leconte2017}
we get $\omega = 0.12 $~eV for $c_{\rm AB}$ 
such that $t_1 = 3 \omega = 0.36$~eV, close to the LDA interlayer coupling in Bernal BG,  
and $\omega^{\prime} = 0.098$~eV for $c_{\rm AA}$ 
which is incidentally rather close to the interlayer tunneling from explicit integrations in $\vec{d}$ in Ref.~[\onlinecite{jung2014}].
The effects of atomic relaxation in the moire patterns can have non-negligible effects in the details 
of the electronic structure for both intralayer potentials and interlayer coupling that can be captured with higher order harmonics
in the moire $G$-vectors.~\cite{jung2015}
It was also noted that in tBG unequal interlayer coupling values $\omega \neq \omega^{\prime}$ enhances the gap between 
the low energy flatband and its neighboring higher energy band.~\cite{koshino_wannier}
The band structures for type I AB/AB structures near $0^{\circ}$ alignment calculated for the minimal tBBG model are shown in Fig.~\ref{figure2} both for 
rigid unrelaxed $\omega = \omega^{\prime}$ and out of plane relaxed interlayer tunneling values $\omega \neq \omega^{\prime}$.
In the case of tBG it was noted that the magic angle follows the $\theta \propto \omega / \upsilon_{\rm F}$ proportionality~\cite{chittari2019} where the 
magic angles grow with increasing interlayer coupling strength. 
Hence, an enhanced $\upsilon_{\rm F} = 1.0\times 10^{6}$~m/s (or $t_0 = -3.1$~eV), 
together with ab initio tunneling $\omega=0.12$~eV, $\omega^{\prime}=0.098$~eV from Eqs.~(\ref{omega}-\ref{omegap})
leads to similar low energy bands as the LDA $\upsilon_{\rm F} = 0.84\times 10^{6}$~m/s (or $t_0 = -2.6$~eV) 
combined with weaker $\omega = 0.1$~eV and $\omega^{\prime}=0.08$~eV values.
In the calculations to follow we use the enhanced Fermi velocity of $\upsilon_{\rm F} = 1.0\times 10^{6}$~m/s together with
the ab initio interlayer tunneling $\omega$ and $\omega^{\prime}$ based on Eqs.~(\ref{omega}-\ref{omegap}) compatible with EXX+RPA equilibrium distances.
Our calculations have used a configuration space with variable cutoff in momentum space of a radius of up to $6 G_1=24\pi \theta/ (\sqrt{3} a)$ 
using Hamiltonian matrices with sizes as large as 676$\times$676 such that $\theta \gtrsim \omega / ( 12 \pi \left| t_0 \right| )$ to obtain converged 
results in the limit of small $\theta$ and large $\omega$.

An important distinctive feature of tBBG with respect to tBG is that we have an additional control knob to change the electronic 
structure through a perpendicular external electric field that modifies the interlayer potential $\Delta_i$ values in Eq.~(\ref{hamiltonian}).
The potential drops introduced by an external electric field could be modeled through the parameter set 
$\Delta_1 = - \Delta_4$, $\Delta_2 = -\Delta_3$, redefined as
$\Delta_1 = ( \Delta + \Delta' )/2$, $\Delta_2 = ( -\Delta + \Delta' )/2$
in terms of $2\Delta$, the interlayer potential difference within each BG, and $2\Delta'$ the potential difference between the BG.
We will use the relation $\Delta^{\prime} = 2\Delta$ to introduce equal interlayer potential drops of $\Delta^{\prime}$ between 
the consecutive layers to model the effects of an electric field. 
A qualitatively different interlayer potential configuration consists in having the electric fields point in opposite directions at each BG. 
This could be done by grounding the tBBG device and using equipotential top/bottom gates to accumulate charges of the same sign at the outer layers. 
The potential distribution of this case can be modeled by $\Delta_1=-\Delta_2=  -\Delta_3 = \Delta_4 = \Delta$,
where the reversal of the relative mass sign between the top and bottom BG can modify the topology of the resulting flatbands.
Distinct band topologies are thus expected for a system subject to a perpendicular electric field
near 180$^{\circ}$ (or $60^{\circ}$) alignment where for the same mass sign the chirality of the bands at $K$ are reversed.

\section{Flatbands as a function of twist angle, electric fields and pressure}
In the following we discuss the electronic structure results and the moire flatband bandwidth in tBBG 
as a function of system parameters such as twist angle $\theta$, the inter-layer coupling $\omega$ tunable by pressure,
and interlayer potential differences due to an electric field, 
in search of the optimal conditions for finding isolated low energy flatbands near the Fermi level.
The interlayer potential differences due to an electric field 
are modeled combining potential differences between the layers within each BG ($\Delta$) and inter-BG potential 
offset ($\Delta^{\prime}$) which are related to each other through $\Delta^{\prime}=2\Delta$
that can lift the degeneracy of the flatbands.

In the band structures resulting from the minimal model shown in Fig.~\ref{figure2} for a twist angle of $\theta=1.5^{\circ}$,
we can still distinguish features of the original BG band structure at the Dirac cones, 
and can identify the band structure near the magic angle $\theta = 1.05^{\circ}$ of tBG.
The first important observation is that the overall bandwidths of the low energy bands in tBBG are almost half 
of those corresponding to tBG for a similar range of $\theta$ and $\omega$ parameter values, suggesting that the 
tBBG system is generally more suitable for the generation of narrow bandwidth flatbands than in tBG.
This is shown in Fig.~\ref{figure3} where we represent the bandwidth as a function of twist angle $\theta$ for 
fixed values of interlayer coupling $\omega$ and as a function of $\omega$ for fixed $\theta$ values. 
From the bandwidth versus $\theta$ dependence we can observe that the bandwidths remain below 10~meV 
for every twist angle below and around the first magic angle. 
Likewise, the bounce off of the bandwidth for increasing $\omega$ past the critical value at the first magic angle 
have maxima that are roughly half of those seen in tBG.~\cite{chittari2019}
We thus expect that in tBBG the twist angle control does not need to be as precise as in tBG 
to maintain a moderately narrow bandwidth on the order of $\sim$10 meV for twist angles smaller than $\sim 1^{\circ}$.
Inclusion of remote hopping terms results in band gaps near charge neutrality for sufficiently large twist angles
and widening of the bandwidths with respect to the minimal model, as shown in Fig.~\ref{figure4remote}.
The trigonal warping term in BG generates several band touching points in the vicinity of the Dirac point, 
and introduces particle-hole symmetry breaking of sufficient relevance especially when we apply an external electric field.
In Fig.~\ref{figure4remote} we compare the band structures of the minimal model and the more complete model that includes
the remote hopping parameters, both for type I near 0$^{\circ}$ and type II near $180^{\circ}$ alignments.
For the complete Hamiltonian we include the remote hopping terms $t_3$ and $t_4$, the site potential offset $\delta$ between the high energy dimer sites,
and interlayer coupling $\omega$ and $\omega^{\prime}$ in Eqs.~(\ref{omega}-$\ref{omegap}$) evaluated at the equilibrium out of plane relaxed lattice distances.

The second important observation is the tunability of the primary and secondary band gaps
accompanied by a variation in bandwidth due to an electric field, as illustrated in Fig.~\ref{figure4remote}.
Even for a twist angle $\sim$1.5$^{\circ}$ considerably greater than the minimal model magic angle of
$\sim 1^{\circ}$ the low energy bands can remain isolated thanks to the primary $\delta_p$ and secondary $\delta_s$ gaps. 
These gap values are obtained from the difference between the maximum (minimum) energy 
of the flatband and the minimum (maximum) energy of the neighboring higher (lower) band 
resulting in positive values when there is a gap and giving negative values when there is band overlap.
Application of external fields contributes in changing the bandwidth and 
a relatively narrow bandwidth on the order of $\sim$25~meV or smaller is achievable for moderate electric fields
that introduce interlayer potential differences typically of a few tens of meV. 
The bandwidths were obtained from the difference between the maximum and minimum band energy for a given band within the mBZ.
Band gaps within each BG layer ($\Delta$) and band offsets between BG layers ($\Delta'$) are simultaneously present when a 
perpendicular electric field is applied in the system. It is found that generally $\Delta^{\prime}$ contributes in widening the bandwidth of the low energy bands.
An important factor for the onset of the interaction driven ordering is the isolation of the low energy bands that 
can be quantified from the primary gap $\delta_{p}$ near charge neutrality and the secondary gap $\delta_{s}$ near the $\Gamma$ 
point of  mBZ, since greater band isolation reduces screening and strengthens the Coulomb interactions.
Hence, the parameter space most likely to observe Coulomb driven ordered phases should have 
simultaneously smaller bandwidths $W$ and larger isolation gaps $\delta_{p}$ and $\delta_{s}$. 
We can estimate the ratio of $U_{\rm eff}/W$ from the effective 3D screened Coulomb potential
\begin{eqnarray}
U_{\rm eff} = \frac{e^2 }{ 4\pi \varepsilon_r \varepsilon_0 l_M } \exp({- l_M / \lambda_{\rm D}})
\label{ueffeq}
\end{eqnarray} 
where the moire length is $l_M = a/\theta$, and the Debye length
$\lambda_{\rm D} = 2 \varepsilon_0/ e^2 D(\delta_p,\delta_s)$ uses the 2D  density of states
$ D(\delta_p,\delta_s) =  4\, \left( \left| \delta_p \right| u(-\delta_p) + \left| \delta_s \right| u(-\delta_s) \right) / (W^2 A_M) $ 
that assumes a value proportional to the band overlap ratio $\delta_{p/s}/W$ when $\delta_{p/s} < 0$,
where $u(x)$ is the heaviside step function, $\varepsilon_r = 4$ and we counted four valley-spin degenerate electrons per moire unit cell area 
$A_M = \sqrt{3} \, l_M^2/2$ for each filled moire band.
This ratio in Eq.~(\ref{ueffeq}) is used to find the parameter space region of twist angle and interlayer electric field with narrow 
bands and strong effective Coulomb interactions, see Fig.~\ref{ratio}. 
While the parameter region near $\theta \sim 0.5^{\circ}$ show largest $U_{\rm eff}/W$ ratios compared to $\theta \gtrsim 1^{\circ}$
due to the greater flatness of the bands at small twist angles, 
the closer proximity of the neighboring energy bands in this regime may enhance the Coulomb screening 
in a way that is not captured in the screening model we have used. 
We expect that the electron-hole asymmetry resulting from the intrinsic asymmetry of BG with remote hopping terms can be further affected
by the coupling with the substrate, for example by aligning BG with a hexagonal boron nitride substrate
which in a heterojunction with single layer graphene it introduced a strong intrinsic particle-hole asymmetry in the electronic structure.~\cite{dasilva-particle-hole}
A phase diagram of the bandwidth, the primary gap, and secondary gap 
as a function of twist angle and external field strength is presented in Fig.~\ref{figure5n}.
For simplicity we have assumed that the potential differences between contiguous layers are given by $\Delta' = 2 \Delta$ 
and are the same for a given electric field, although the precise interlayer potential differences will depend on the screening between the layers.  
This phase diagram of $U_{\rm eff}/W$ ratios, bandwidths, primary, and secondary gaps both for electrons and holes illustrates the parameter
space where the likelihood for finding ordered phases is higher. 
As a general trend we find a non-negligible asymmetry between electrons and holes, 
and the possibility of finding states with primary and secondary gaps both at small and large twist angles of 
$\sim 0.5^{\circ}$ and $\sim 1.5^{\circ}$ for sufficiently strong electric fields. 
Due to the almost linear increase of the low energy bandwidth with increasing twist angle 
there is a tradeoff between band isolation easier at larger twist angles and bandwidth increase to find the optimum parameter space where the Coulomb interaction effects will be strongest.

Application of pressure is also a useful control knob to tune the electronic structure of 2D materials.~\cite{yankowitz1,Yankowitz,kaxiras2018,chittari2019}
In the bandwidth and gaps phase diagram as a function of twist angle and pressure shown in Fig.~\ref{figure_pressure}
both for $\Delta^{(\prime)} =0 $ and  $\Delta^{(\prime)} \neq 0 $ we can observe that the enhancement of $\omega$
through pressure generally compresses the bandwidth of the flatbands.
In the minimal model of tBBG the application of pressure leads to a magic angle line
\begin{eqnarray} 
\theta^{\circ}_{n} &=&  C_n \dfrac{\omega}{|t_{0}|} \quad  ({\rm deg})
\label{mangle}
\end{eqnarray}
whose coefficient values of $C_1 = 27.5$, $C_2 = 10.5$, and $C_3 = 5.6$ 
agree within~10\% with the results obtained for tBG,~\cite{chittari2019}
in keeping with the expected band scaling behavior proportional to $\alpha = \omega/(\theta \upsilon_{\rm F} K_D)$,~\cite{bistritzer2011}
see supplemental materials.~\cite{supplemental}   
This observation suggests that increase of $\omega$, e.g.  through external pressure $P$,~\cite{yankowitz1}
should allow to achieve narrow bandwidth features for larger twist angle $\theta$.
This behavior holds both for the minimal model and when remote hopping terms are considered,
although the remote hopping terms tend to broaden the bandwidth of the minimal model, 
see the supplemental information for a comparison of the phase diagrams.~\cite{supplemental} 
The phase diagram also illustrates how pressure can be used to enhance 
the secondary gaps both for valence and conduction bands, 
but it can at the same time suppress the primary gap. 
The phase diagram is modified substantially in the presence of interlayer potential differences $\Delta^{(\prime)} \neq 0$
triggered by a perpendicular electric field, generally widening the bandwidth in the phase diagram
and shifting the weights between the primary and secondary gaps. 
The relationship between $P$ and $\omega$ values in the relevant range of pressures between 
0 to $\sim$15~GPa if fitted by a second order polynomial and its positive root
\begin{figure}
\begin{center}
\includegraphics[width=8.2cm]{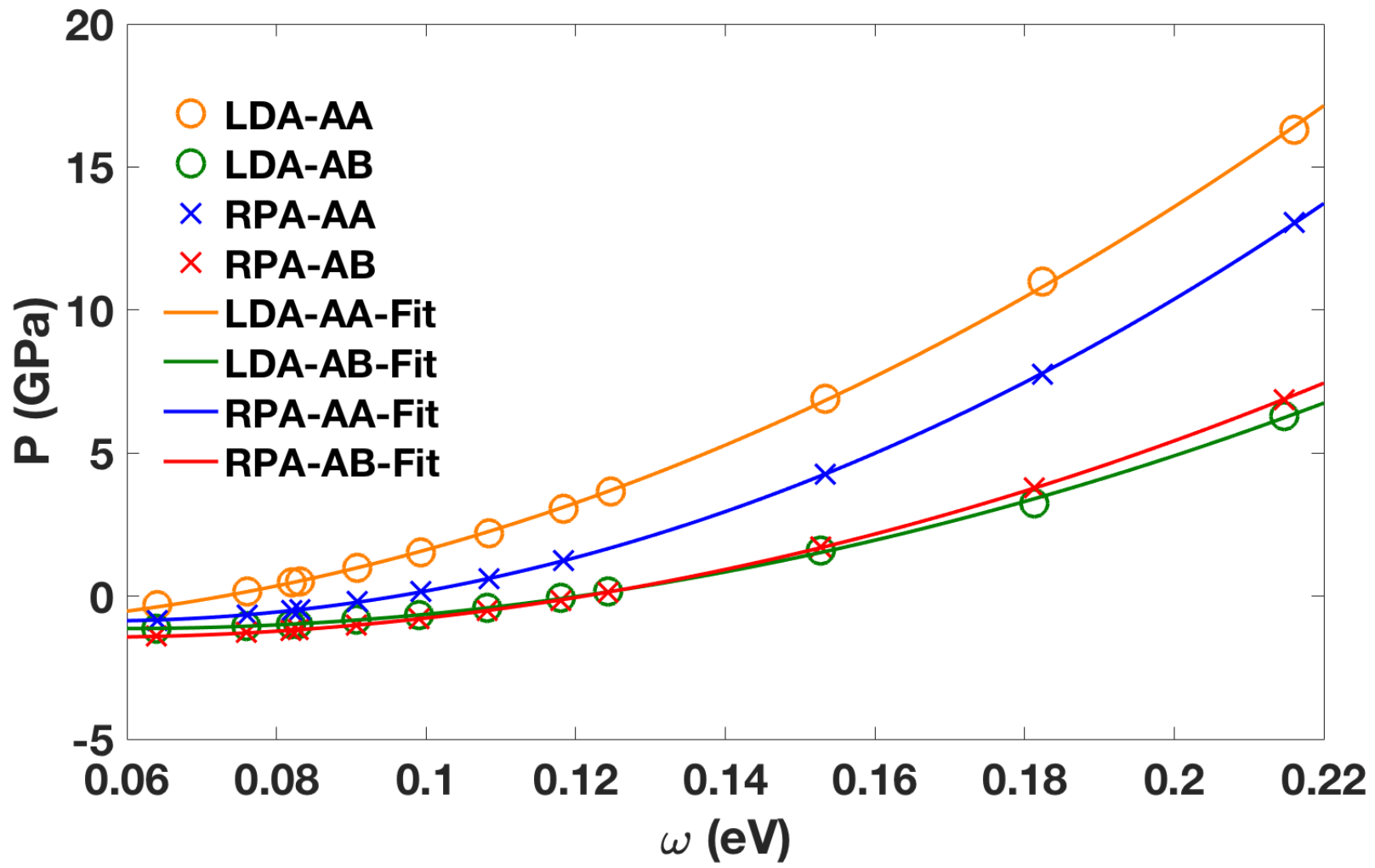}
\end{center}
\caption{
(Color online) Interlayer coupling versus pressure for different stacking geometries AA, AB (equivalent to BA) 
for equilibrium distances calculated within the LDA and EXX+RPA calculations from Ref.~[\onlinecite{leconte2017}]
that can be fitted through Eqs.~(\ref{pvsomegaeq}-\ref{omegavspeq}) with the parameters in Table~\ref{parameters}. 
}
\label{pvsomega}
\end{figure}
\begin{eqnarray}
P &=& A \omega^2 + B \omega + C   \label{pvsomegaeq} \\
\omega &=& A^{\prime} + \sqrt{ B^{\prime} + C^{\prime} P }. \label{omegavspeq}
\end{eqnarray}
\begin{table}
\caption{
External pressure $P(\omega_{s})$ and associated interlayer 
coupling $\omega_s(P)$ for commensurate stacking geometries ($s$ = AA, AB, BA) illustrated in Fig.~\ref{pvsomega}.
We list the fitting coefficients for $P(\omega)$ in Eq.~(\ref{pvsomegaeq}) $A$ ({GPa}/eV$^{2}$), 
for $B$ ({GPa}/{eV}), and $C$ (GPa),
and the inverse fit for $\omega(P)$ in Eq.~(\ref{omegavspeq}) 
$A^{\prime}$ (eV), for $B^{\prime}$ (eV$^{2}$), and $C^{\prime}$ (eV$^2$/GPa).
}
\begin{center}
\begin{tabular}{cc|c|c|c|c|c|c}\\ \hline \hline 
                        &&     \multicolumn{3}{c|}{LDA}  &  \multicolumn{3}{c}{EXX+RPA}    \\ \hline \hline
Stacking   (s)   &&     \multicolumn{1}{c|}{  $A$     }          &  \multicolumn{1}{c|}{  $B$  }       &    \multicolumn{1}{c|}{  $C$    }
                          &     \multicolumn{1}{c|}{  $A$     }          &  \multicolumn{1}{c|}{  $B$  }       &    \multicolumn{1}{c}{  $C$    }   \\  \hline
AA               &&   473.6 & -22.17  & -0.9011  & 543.3  & -61.01 & 0.735    \\ \hline
AB               &&   306.5 & -36.56  &  -0.04339 & 324.7 & -35.47& -0.4671     \\  \hline \hline
Stacking   (s)   &&     \multicolumn{1}{c|}{  $A^{\prime}$     }          &  \multicolumn{1}{c|}{  $B^{\prime}$  }       &    \multicolumn{1}{c|}{  $C^{\prime}$    }
                         &     \multicolumn{1}{c|}{   $A^{\prime}$     }           &  \multicolumn{1}{c|}{  $B^{\prime}$  }       &    \multicolumn{1}{c}{  $C^{\prime}$    }   \\  \hline
AA               &&   0.0234 & 0.0025  & 0.0021  & 0.0561 & 0.0018 & 0.0018    \\ \hline
AB               &&   0.0596 & 0.0037  &  0.0033  & 0.0546 & 0.0044 & 0.0031     \\  \hline \hline
\end{tabular}
\label{parameters}
\end{center}
\end{table}
The fitting parameters are listed in Table~\ref{parameters} and they are found to be valid over a wide range of pressures stretching up to $\sim$30~GPa and also for negative values down to $\sim-1$~GPa.
To obtain the above fitting parameters we have used the relationship between $P$ and interlayer separation $c$ for every stacking of Ref.~[\onlinecite{leconte2017}], 
and the calculations of $\omega$ versus $c$ as detailed in Ref.~[\onlinecite{jung2014}]. The explicit fitting functions for these quantities are presented in the supplemental material.~\cite{supplemental} 
This approach is different to that in Ref.~[\onlinecite{chittari2019}] where the total pressure was obtained from the
average of the local pressure values at the same interlayer distance $c$ for every stacking.

\begin{figure*}
\begin{center}
\includegraphics[width=17.5cm]{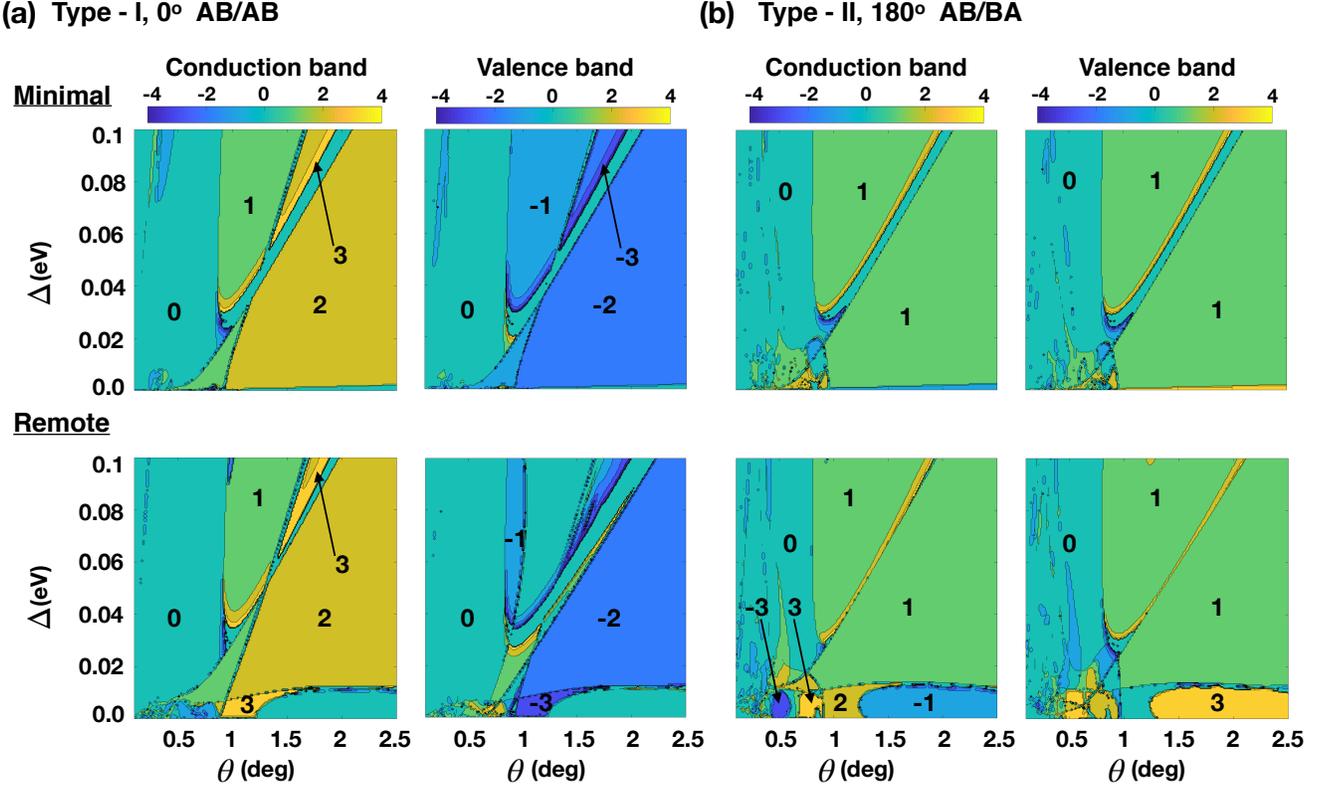}
\end{center}
\caption{
(Color online)
The $K$ valley Chern number phase diagram of low energy conduction and valence bands 
for the minimal model of BG (top row) and including the remote hoping parameters (bottom row).
Agreement between both models are seen for a large parameter space,
although when remote hopping terms are considered we can  
trigger quantum phase transitions with different valley Chern numbers at moderate values of $\Delta \sim 10$~meV to 
access regions in phase space where we can expect quantized Hall conductivities at zero magnetic fields,
particularly for twist angles $\theta \gtrsim 1^{\circ}$. 
}
\label{chernnumbers}
\end{figure*}

\begin{figure*}
\begin{center}
\includegraphics[width=18cm]{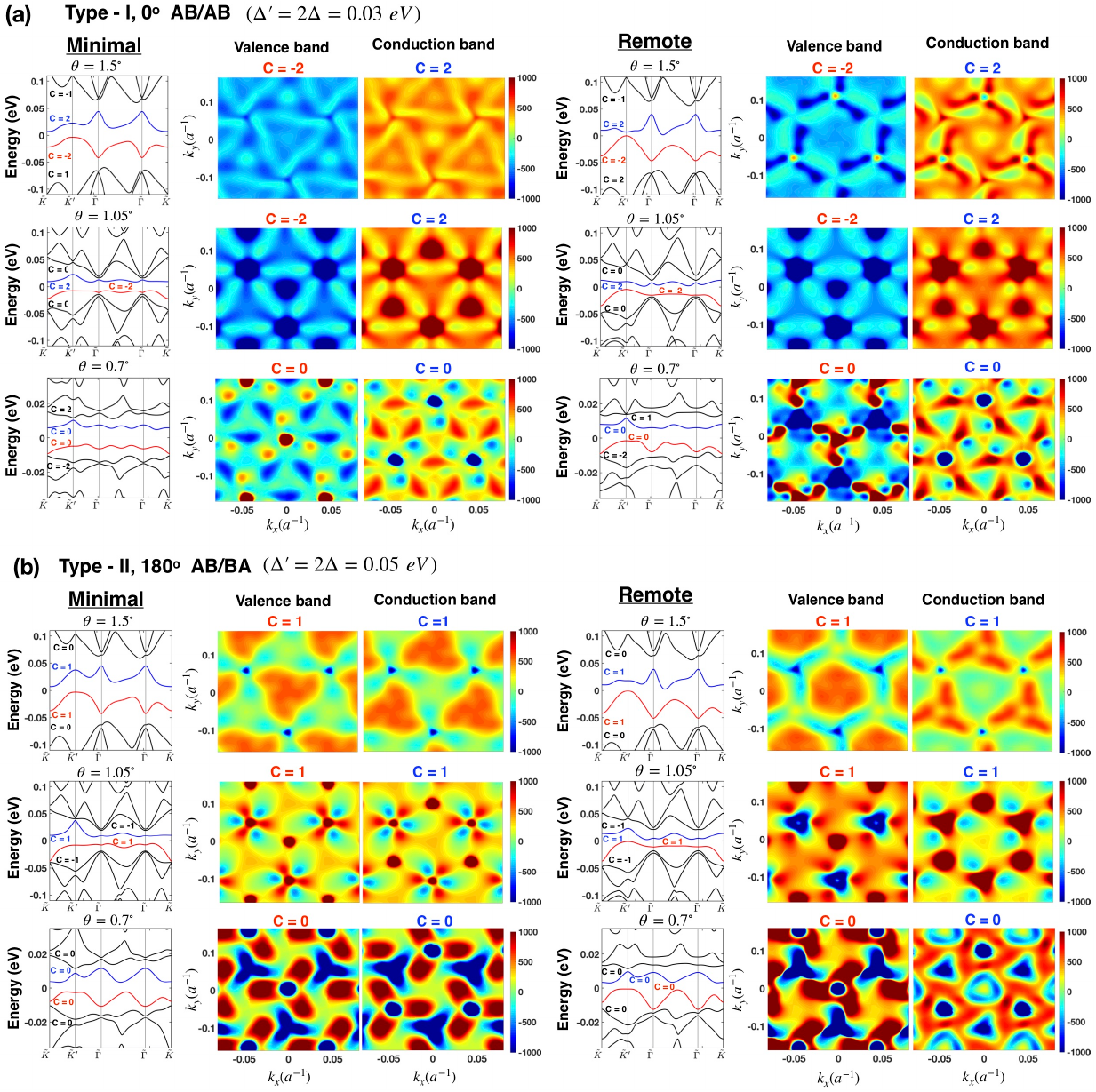}
\end{center}
\caption{
(Color online) Band structure and Berry curvature plots for select twist angles $\theta = 0.7^{\circ}, 1.05^{\circ}, 1.5^{\circ}$ and interlayer 
potential differences within the minimal and remote hopping parameters Hamiltonian model for BG,
leading to quantitative differences in the bandwidths and associated $\delta_{p/s}$ for the primary and secondary gaps.
The results are shown for (a) type-I near $0^{\circ}$ AB/AB alignment for $\Delta^{\prime} = 2 \Delta = 0.03$~eV 
and for (b) type-II near $180^{\circ}$ AB/BA alignment that requires a larger $\Delta^{\prime} = 2 \Delta = 0.05$~eV to 
simultaneously achieve positive primary and secondary gaps $\delta_{p/s} >0$. 
For $\theta \gtrsim 1^{\circ}$ these two alignments are shown to give distinct $C_{\upsilon} = \pm 2$ of opposite and $C_{\upsilon} =  1$ same sign
$K$ valley Chern numbers for valence and conduction bands. These qualitative differences in the Chern numbers for the two
stacking alignments can be traced to the different associated chiralities of the bands in the top and bottom layers.
}
\label{BC}
\end{figure*}

\section{Topological flatbands in twisted bi-bilayer graphene}
The presence of moir\'e superlattices give rise to avoided gaps between the bands in the moire mini Brilloun zone (mBZ) 
allowing them to have a well defined valley Chern number in a wide class of twisted multilayer systems,~\cite{todadri,senthil1,zaletel}
transition metal dichalcogenides,~\cite{wumacdonald}
trilayer graphene on hexagonal boron nitride (TG/BN),~\cite{prljung}
and for a variety of twisted gapped Dirac materials.~\cite{javvaji}
For our tBBG system the possibility of opening a band gap $\delta_{p}$ near charge neutrality through an electric field
together with the opening of a secondary gap $\delta_{s}$ with the higher energy bands 
makes tBBG an interesting platform to engineer flatbands with well defined valley Chern numbers that are tunable 
through an electric field like in TG/BN.~\cite{todadri,prljung} 
The valley Chern number phase diagram in Fig.~\ref{chernnumbers} as a function of interlayer potential differences and twist 
angle indicates the range of Chern numbers $C_{\upsilon} = 0,\, \pm1,\, \pm2,\, \pm3,\, \pm4$ expected in these systems.
The valley Chern numbers were calculated through 
\begin{eqnarray}
C_{\upsilon} = \int_{\rm mBZ} {\rm d^2} \vec{k} \,\, \Omega_{n}(\vec{k})/(2\pi)
\end{eqnarray} 
by integrating in the moire Brillouin zone for each valley the Berry curvature for the $n^{th}$ band through~\cite{rmp_berry}

\begin{eqnarray}
\Omega_n (\vec{k}) = -2 \sum_{n' \neq n} {\rm Im} \left[  \frac{ \langle u_n | \frac{\partial H}{ \partial k_x} | u_{n'} \rangle    \langle u_{n'} |  \frac{\partial H}{ \partial k_y} | u_{n} \rangle 
}{ \left( E_{n'} - E_n \right)^2 } \right]
\end{eqnarray}
where for every $k$-point we take sums through all the neighboring $n'$ bands, 
the $| u_n \rangle$ are the moir\'e superlattice Bloch states and $E_n$ are the eigenvalues.
There are clear qualitative differences between the valley Chern numbers for twist angles near $0^{\circ}$ for type-I AB/AB and those near $180^{\circ}$ for type-II AB/BA alignments. 
In the first case the valley Chern numbers between valence and conduction bands are generally opposite in value adding up to a total of zero,
while in the second case they are the same number for both valence and conduction bands. 
These differences are naturally expected if we consider that the chirality of the massive bands at the top and bottom layers that couple to each other 
are interchanged depending on the alignment.  
Quantitative modifications in the valley Chern numbers are observed when we compare the phase diagrams of the minimal
and remote hopping parameter models as a function of $\Delta$ where clear differences in particular for
small $\Delta \sim 10$~meV region is observed,
comparable in magnitude with the band distortions introduced by the remote hopping terms. 
The vicinity of $\theta \sim 1^{\circ}$ and low electric field $\Delta$ have spots where the  
valley Chern numbers for the valence and conduction bands differ in magnitude.
For sufficiently large values of $\Delta$ the valley Chern numbers of the lowest energy flatbands
agree for the minimal and remote hopping parameter models in a large parameter space of twist angles
as illustrated in Fig.~\ref{BC} by showing the band structure and the Berry curvatures used in the valley Chern number calculations.
Hence, the topological properties of the bands remain overall relatively robust to small perturbations 
to the band structures introduced by the remote hopping parameters. 
The values of the low energy flatband Chern numbers and those of the neighboring higher energy bands 
are gathered in Table~\ref{ChernValues}.  
When the valley Chern numbers between the minimal and remote hopping term models 
differ are comparable in a large parameter space.

\begin{table}
\caption{The Chern numbers table of tBBG for twist angles near the 0$^{\circ}$ (Type-I) and 180$^{\circ}$ (Type-II) 
alignment with and without remote hoping parameters 
at the angles $\theta = 1.5^{\circ}, 1.05^{\circ}$ and 0.7$^{\circ}$ whose band structures are represented in Fig.~\ref{BC}. 
The Chern numbers presented here are for low energy conduction (C) and valence bands (V), 
and one higher energy bands in each conduction (C$+$1) and valence (V$-$1) bands. 
The interlayer potentials are respectively $\Delta^{\prime}=0.03$~eV 
for $\theta \sim 0^{\circ}$ and $\Delta^{\prime}=0.05$~eV for $\theta \sim 180^{\circ}$
that are large enough to isolate the low energy bands.
}
\begin{center}
\begin{small}
\begin{tabular}{cc|c|c|c|c|c|c|c|c}\\ \hline 
                     && \multicolumn{4}{c|}{Minimal}& \multicolumn{4}{c}{Remote } \\ \hline  
Bands          &&\multicolumn{1}{c|}{V ($-$1)}& \multicolumn{1}{c|}{V }& \multicolumn{1}{c|}{ C}& \multicolumn{1}{c|}{C ($+$1)} 
                       &\multicolumn{1}{c|}{V ($-$1)}& \multicolumn{1}{c|}{V }& \multicolumn{1}{c|}{ C}& \multicolumn{1}{c}{C ($+$1)} \\ \hline
                     && \multicolumn{8}{c}{\bf $\theta = 1.5^{\circ}$}   \\ \hline
Type-I      && 1 & $-$2 & 2  & $-$1 & 2 & $-$2 & 2 & $-$1\\ 
Type-II          && 0& 1  &  1  & 0 & 0 & 1 & 1 & 0\\  \hline 
                     && \multicolumn{8}{c}{\bf $\theta = 1.05^{\circ}$}   \\ \hline
Type-I        && 0 & $-$2  & 2  & 0 & 0 & $-$2 & 2 & 0\\ 
Type-II       && $-$1 & 1  &  1  & $-$1 & $-$1 & 1 & 1 & $-$1\\  \hline 
                     && \multicolumn{8}{c}{\bf $\theta = 0.7^{\circ}$}   \\ \hline
Type-I       && $-$2 & 0  & 0  & 2 & $-$2 & 0 & 0 & $\hdots$\\ 
Type-II        && 0 & 0  &  0  & 0 & 0 & 0 & 0 & 0\\  \hline 
\end{tabular}  \label{ChernValues}
\end{small}
\end{center}
\end{table}

\section{Summary and conclusions}
We have extended the Bistritzer-MacDonald continuum model of twisted bilayer graphene (tBG)
to investigate the electronic structure of twisted bi-bilayer graphene (tBBG)
as a function of twist angle $\theta$,  electric fields, and the pressure dependent inter-layer coupling. 
We have considered both the minimal model and also the effects of the 
remote hopping terms in the band structure calculation of bilayer graphene. 
The calculated bandwidth phase diagram for the low energy bands shows that the bandwiths are
roughly a factor two narrower than those in twisted bilayer graphene (tBG) indicating that  
tBBG should be more forgiving in the twist angle precision required to access the strongly interacting regime,
and for this reason we expect that the narrow band features in tBBG will be observed more simply than in tBG. 
%
%
The possibility of applying a perpendicular electric field is an interesting control knob that allows to enhance the 
separation between the flatbands and also influences the gaps with the higher energy bands favoring a more effective band isolation.
At the same time, we find that interlayer potential differences can widen the bandwidths near the first magic angle of the minimal model
and smoothen the bandwidth variation to give a continuous range of angles where the bandwidths are narrow.
Within the minimal model, the bandwidth phase diagram for zero interlayer potential difference and small perturbations thereof 
is found to be closely similar to the case of tBG, maintaining the same linear dependence between $\theta$ and 
the interlayer coupling $\omega$ for the magic angles, and the inverse proportionality to the Fermi velocity of the graphene layers. 
Our calculations show that bi-bilayer graphene under a perpendicular electric field can host robust ordered phases for twist angles 
in the vicinity of $\sim0.6^{\circ}$ and near $\sim1.5^{\circ}$, with the parameter space for the conduction bands being generally favored over those of the valence bands.
With proper electric fields we expect that practically all angles spanning the range between $0.4^{\circ} - 1.6^{\circ}$ could host ordered phases.
Application of pressure can also enhance the isolation of the bands when used in combination with appropriate electric fields. 
We have related the interlayer tunneling with external pressure through stacking dependent interlayer coupling parameters $\omega$ and $\omega^{\prime}$ compatible with 
the EXX+RPA interlayer potentials to capture the interdependence of corrugation and interlayer tunneling in a self-contained manner. 
A more detailed study that combines the effects in-plane moire strains will be addressed elsewhere.
Comparison between electronic structure between the type I near $0^{\circ}$ aligned systems and type II $180^{\circ}$ aligned bi-bilayers
indicate that $0^{\circ}$ aligned systems generally require weaker electric fields to achieve optimal flatband systems prone to interaction 
driven ordered phases, and give rise to distinct valley Chern number phase diagrams.

\bigskip

\begin{acknowledgments}
We acknowledge helpful discussions with S. Kahn, F. Wu and M. Koshino.
This work was supoprted by Samsung Science and Technology Foundation under project no. SSTF-BA1802-06 for J.J.
Financial support is acknowledged for N.R.C. from the Korean National Science Foundation through grant number NRF-2016R1A2B4010105,
for B.L.C. by the Basic Science Research Program through the National Research Foundation of Korea (NRF) funded by the Ministry of 
Education (2018R1A6A1A06024977) and NRF-2017R1D1A1B03035932.
This research was supported in part by the US National Science Foundation under Grant No. NSF PHY-1748958.
\end{acknowledgments}
{\em Note added}.$-$ Related experiments and theory are Refs. [\onlinecite{philipkim,guangyu,pjarillo}]
and Refs. [\onlinecite{choi,leey,koshi,liu}].

\section{Supplemental material}
\begin{widetext}
Here we provide additional figures to supplement the information in the main text for (1) the pressure dependent bandwidth and gaps phase diagram, (2) the parametrization of the interlayer distance dependent pressure and tunneling for the $P$ versus $\omega$ relationship, and (3) the low energy bands formulation of the Hamiltonian.

\subsection{Effect of interlayer coupling strength }
 The pressure dependent phase diagrams for valence (v), conduction (c) flatband bandwidth and primary $\delta_p$, 
secondary $\delta_s$ gaps between the low energy flatband with the surrounding bands with the variation of interlayer potential difference $\Delta^{\prime}$ for the 0$^{\circ}$ (Type-I) and 60$^{\circ}$ (Type-II) aligned BG/BG are shown to understand the effect of  interlayer potential difference.  We discuss the results obtained within the minimal model for two types of stacking in Fig.~\ref{minimal-1} and Fig.~\ref{minimal-2}.
The effects of remote hopping terms in the Hamiltonian are discussed in Fig.~\ref{remote-1} and ~\ref{remote-2}

%

\begin{figure*}[h]
\begin{center}
\includegraphics[width=17cm]{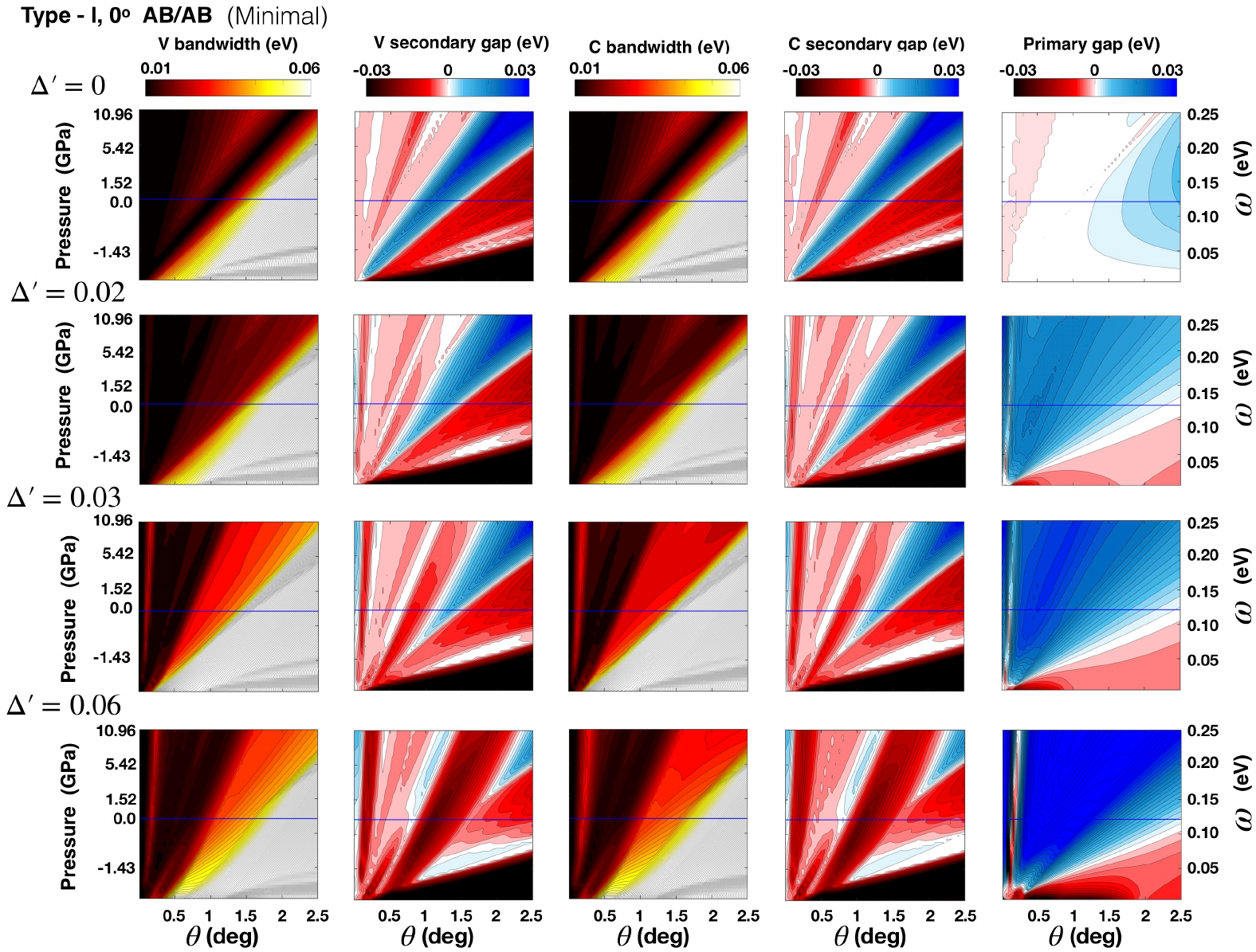}
\end{center}
\caption{
(Color online)
Pressure dependent phase diagram of the valence (v), conduction (c) flatband bandwidth and primary $\delta_p$, 
secondary $\delta_s$ gaps between the low energy flatband with the surrounding bands with the variation of interlayer potential difference $\Delta^{\prime}=~$0, 0.02, 0.03 and 0.06 ~eV for the 0$^{\circ}$ (Type-I) aligned BG/BG. 
The effect of remote hoping terms is absent i.e minimal model on the bandwidth phase diagram is shown, the horizontal blue line which indicates the pressures zero (P = 0 GPa).
}
\label{minimal-1}
\end{figure*}

\begin{figure*}[h]
\begin{center}
\includegraphics[width=17cm]{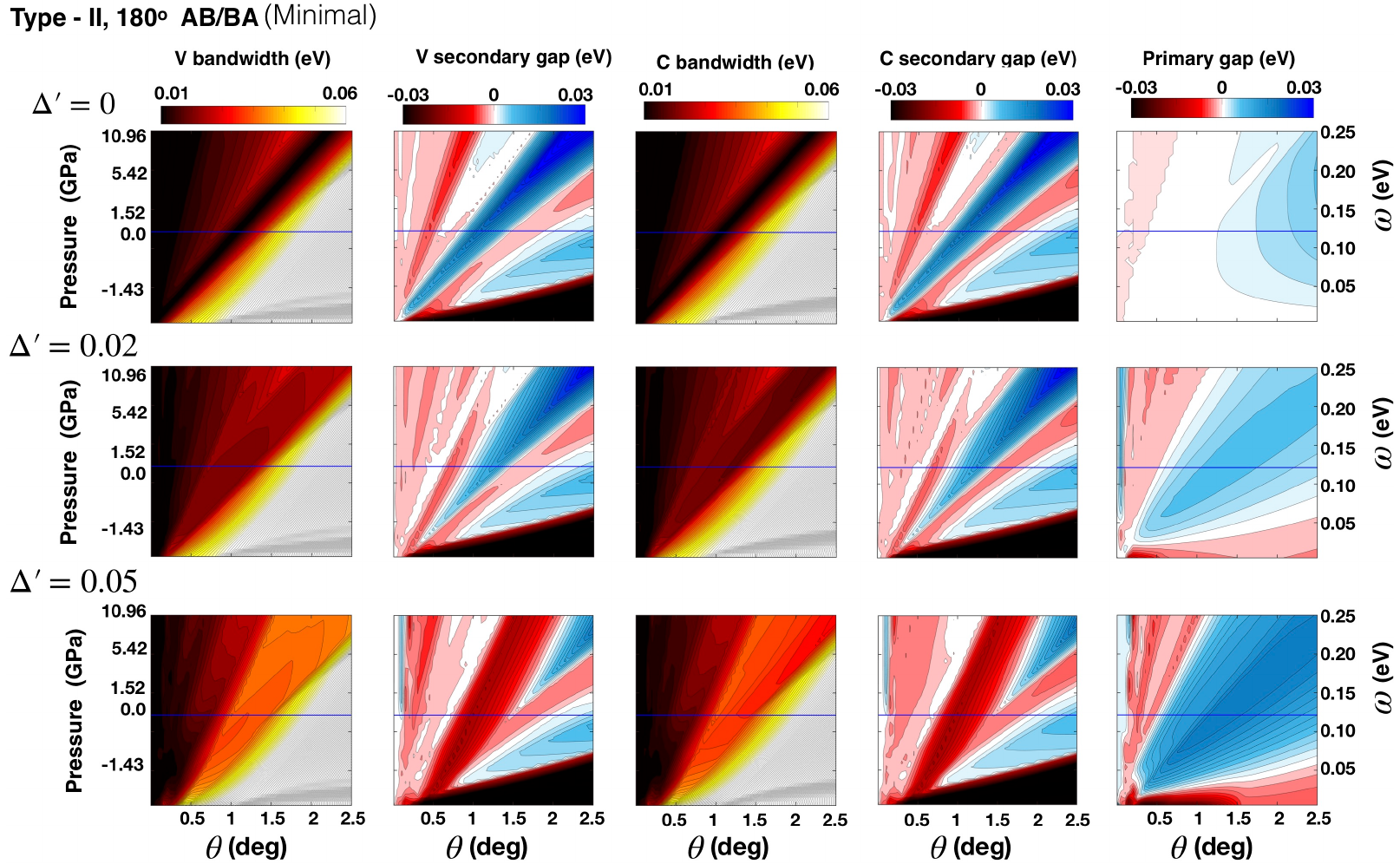}
\end{center}
\caption{
(Color online)
Pressure dependent phase diagram of the valence (v), conduction (c) flatband bandwidth and primary $\delta_p$, 
secondary $\delta_s$ gaps between the low energy flatband with the surrounding bands with the variation of interlayer potential difference $\Delta^{\prime}=~$0, 0.02, and 0.05 ~eV~for the 60$^{\circ}$ (Type-II) aligned BG/BG. 
The effect of remote hoping terms is on the bandwidth phase diagram is shown, the horizontal blue line which indicates the pressures zero (P = 0 GPa).
}\label{minimal-2}
\end{figure*}

\begin{figure*}[h]
\begin{center}
\includegraphics[width=17cm]{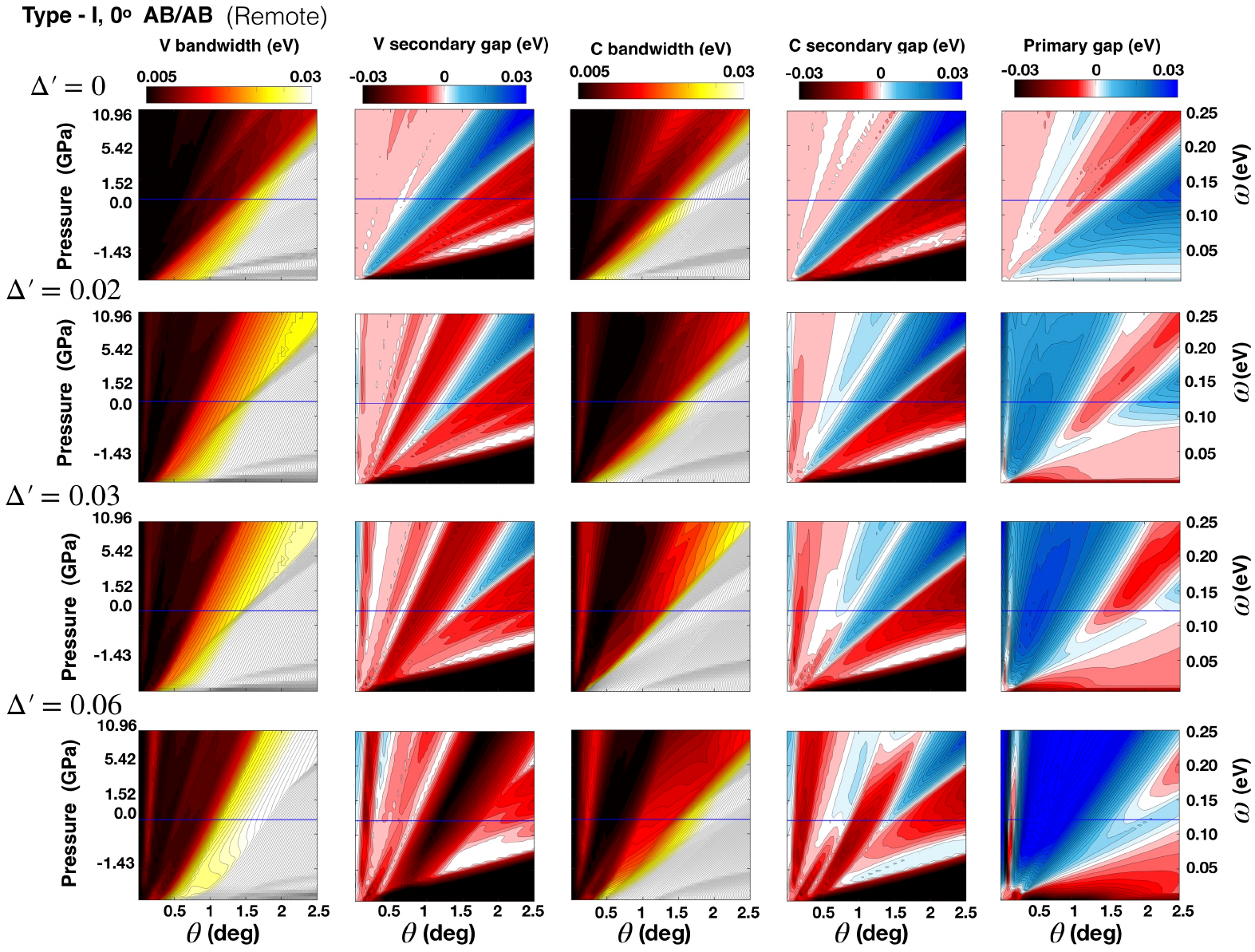}
\end{center}
\caption{
(Color online)
Pressure dependent phase diagram of the valence (v), conduction (c) flatband bandwidth and primary $\delta_p$, 
secondary $\delta_s$ gaps between the low energy flatband with the surrounding bands with the variation of interlayer potential difference $\Delta^{\prime}=~$0, 0.02, 0.03 and 0.06 ~eV for the 0$^{\circ}$ (Type-I) aligned BG/BG. 
The effect of remote hoping terms is absent i.e minimal model on the bandwidth phase diagram is shown, the horizontal blue line which indicates the pressures zero (P = 0 GPa).
}\label{remote-1}

\end{figure*}

\begin{figure*}[h]
\begin{center}
\includegraphics[width=17cm]{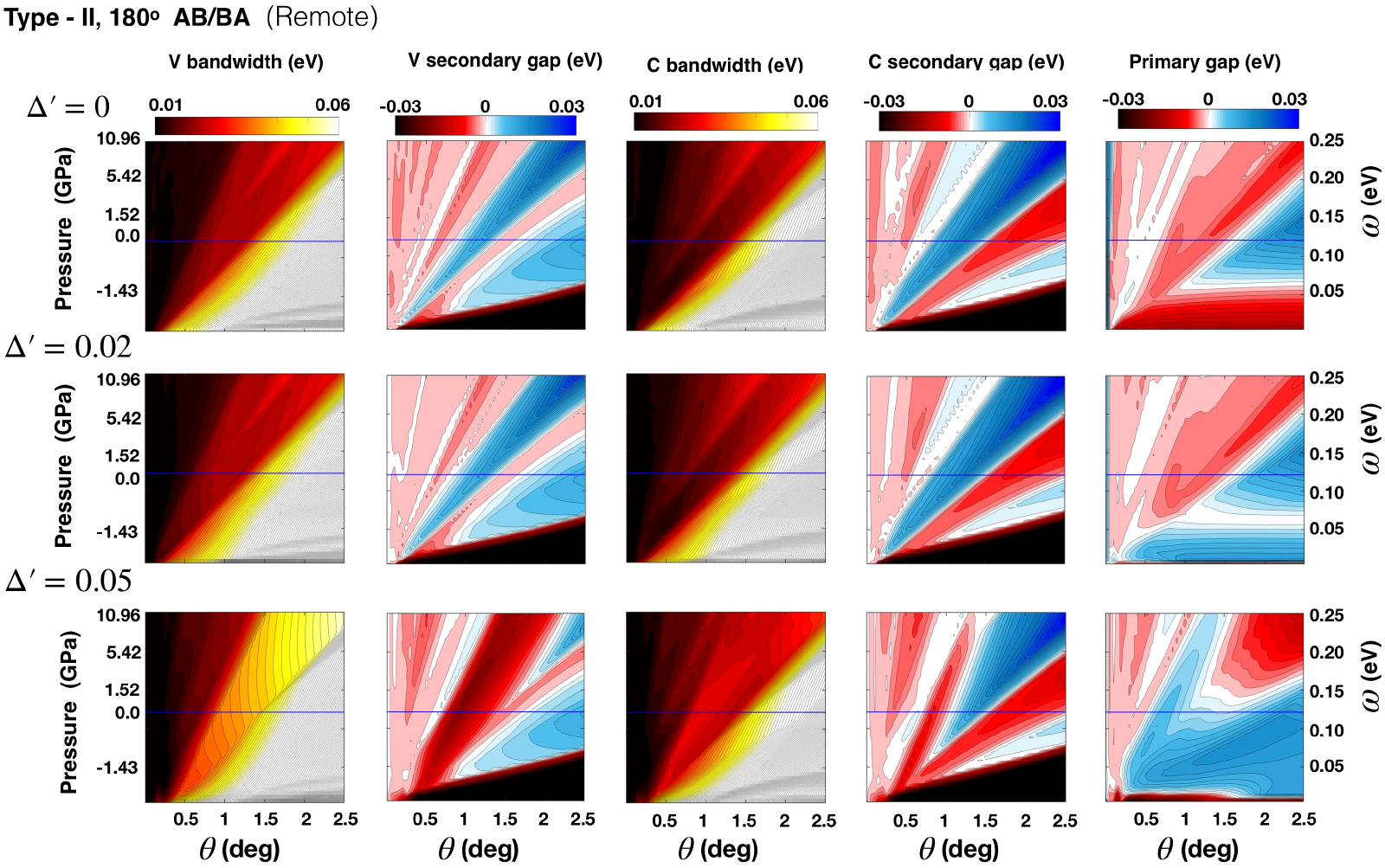}
\end{center}
\caption{
(Color online)
Pressure dependent phase diagram of the valence (v), conduction (c) flatband bandwidth and primary $\delta_p$, 
secondary $\delta_s$ gaps between the low energy flatband with the surrounding bands with the variation of interlayer potential difference $\Delta^{\prime}=~$0, 0.02, and 0.05 ~eV~for the 60$^{\circ}$ (Type-II) aligned BG/BG. 
The effect of remote hoping terms is absent i.e minimal model on the bandwidth phase diagram is shown, the horizontal blue line which indicates the pressures zero (P = 0 GPa).
}\label{remote-2}
\end{figure*}

\clearpage
\subsection{Coulomb interaction potential}
Coulomb interaction driven instabilities arise in tBBG due to narrow bandwidth. The strong correlations are evident from the $U/W$ ratio, and here we take ratio of  
the effective coulomb interaction potential without screening:
\begin{eqnarray}
U = \frac{e^2 }{ 4\pi \varepsilon_r \varepsilon_0 l_M } 
\end{eqnarray}
The Fig.~\ref{ratio} is obtained for the parameter space of $\theta$ and  $\Delta$ .

\begin{figure}
\begin{center}
\includegraphics[width=16.5cm]{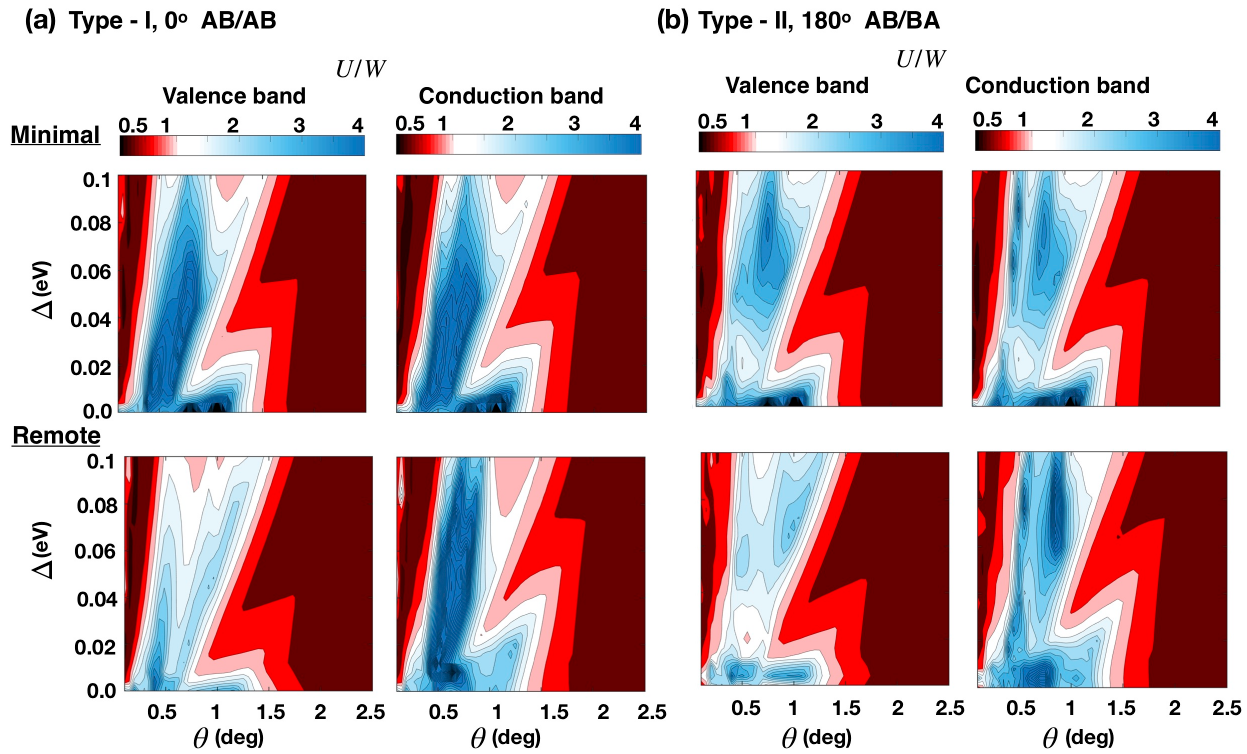}
\end{center}
\caption{
(Color online). The white and blue regions in the colormap phase diagram of $U/W \gtrsim 1$ for  in the colormap in the parameter space of $\theta$ and  $\Delta$ that indicates the plausible regions where Coulomb interactions can play a dominant role.  Screening of Coulomb interactions will be suppressed when $\delta_{p/s} \neq 0$.
}
\label{ratio}
\end{figure}

\clearpage
\subsection{P vs $\omega$ relations}
Here we present the various relations that are using in the main-text, Pressure versus interlayer distance is shown in Fig.~\ref{PC-relation},  $\omega$ versus interlayer distance is given in Fig.~\ref{WC-relation}. The relationship between $\omega^{\prime}$  and  $\omega$ is shown in Fig.~\ref{WC-relation2} for RPA, and in Fig.~\ref{WC-relation3} is for the LDA.

\begin{figure*}[h]
\begin{center}
\includegraphics[width=10cm]{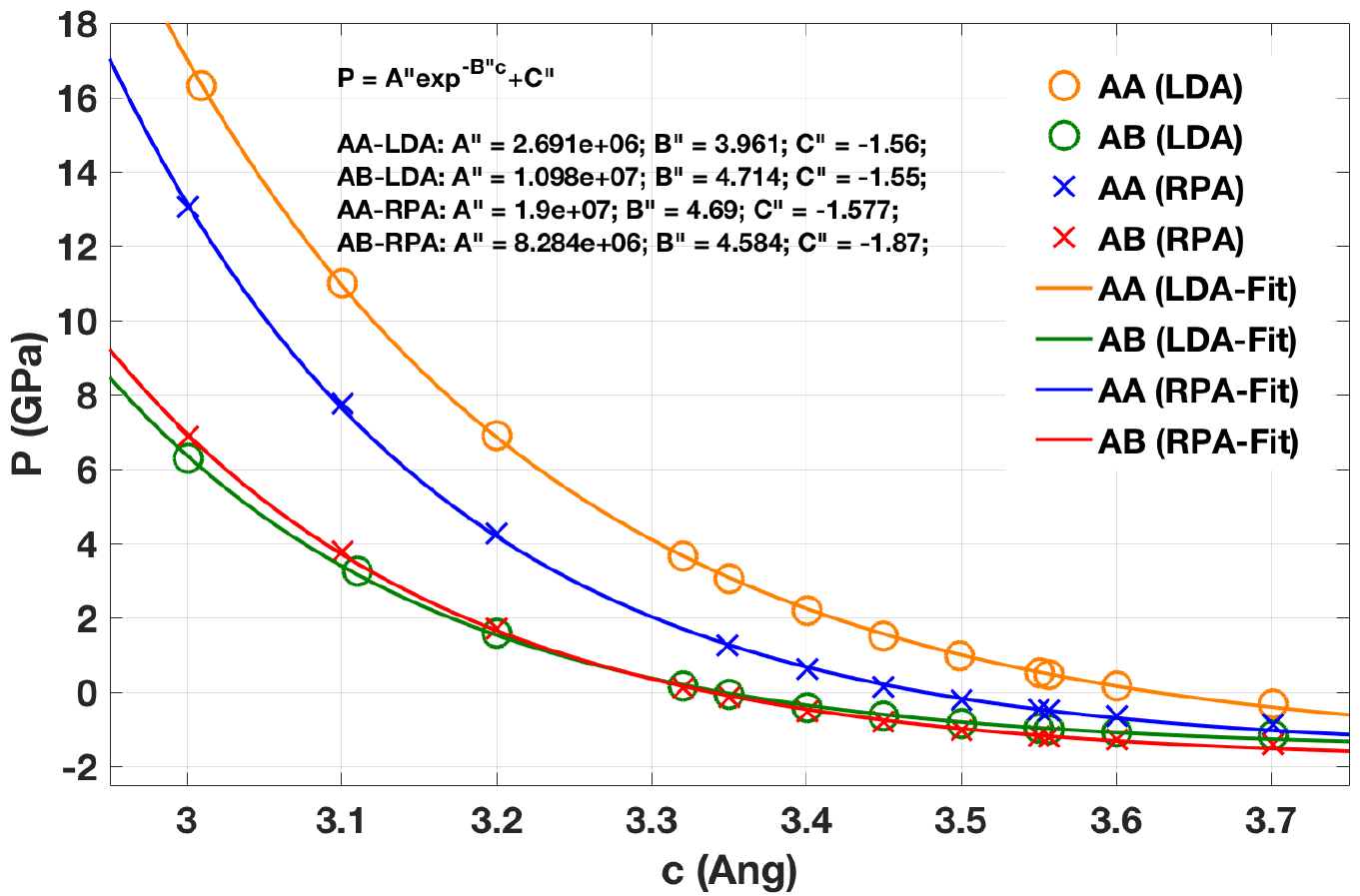}
\end{center}
\caption{
(Color online)
Interlayer distance (c, $\AA$) versus pressure (P, GPa) for different stacking geometries AA, AB (equivalent to BA) 
for equilibrium distances calculated within the LDA and EXX+RPA calculations. The relation between the pressure (P) and interlayer distance (c) is connected through the equation $P = A^{\prime\prime}exp^{-B^{\prime\prime}c}+C^{\prime\prime}$. 
}\label{PC-relation}
\end{figure*}

\begin{figure*}[h]
\begin{center}
\includegraphics[width=10cm]{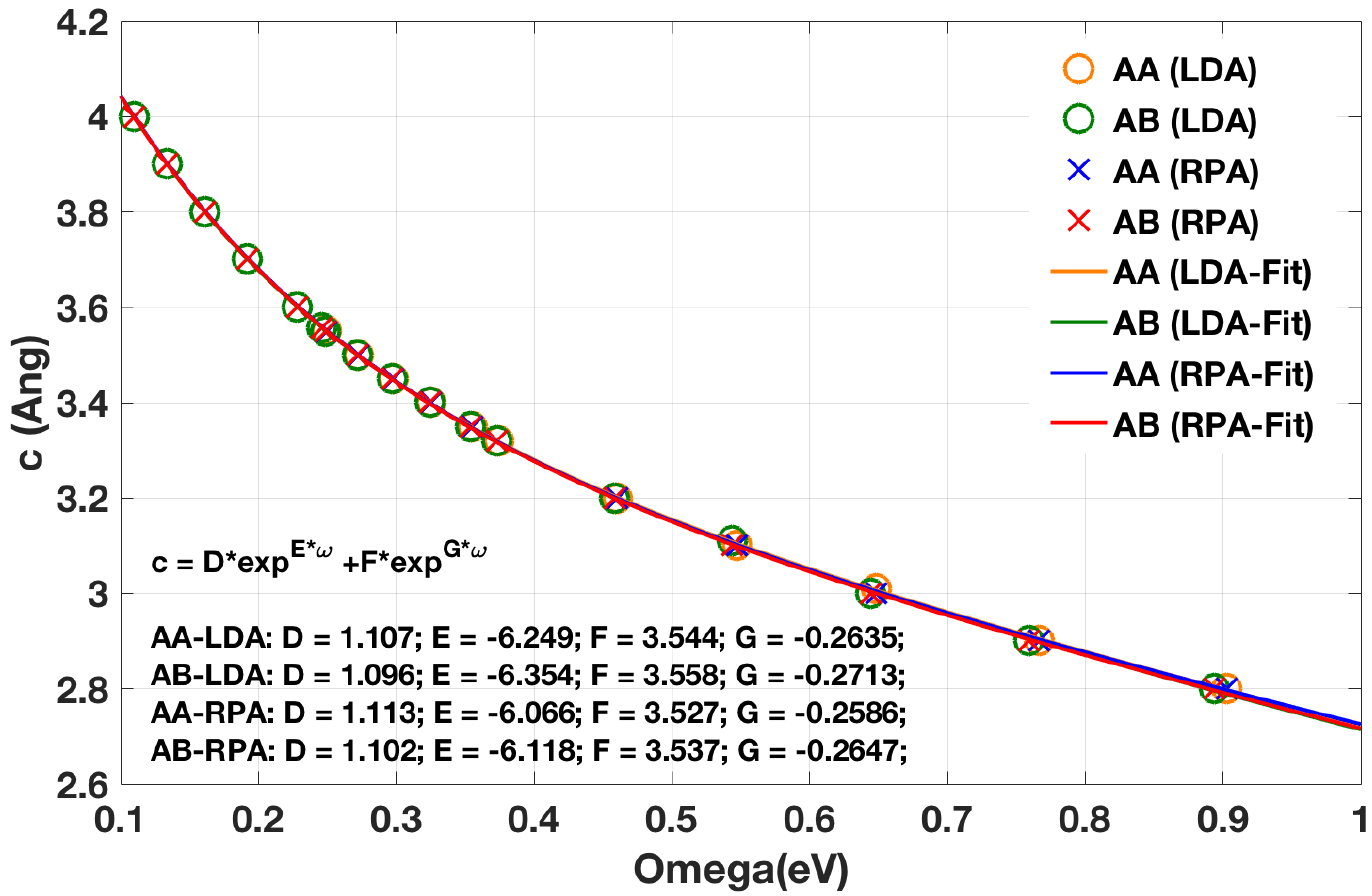}
\end{center}
\caption{
(Color online)
Interlayer coupling strength ($\omega$, eV) versus interlayer distance (c, $\AA$) for different stacking geometries AA, AB (equivalent to BA) for equilibrium distances calculated within the LDA and EXX+RPA calculations. The relation between the $\omega$ and interlayer distance (c) is connected through the equation $c = D \exp({E\omega})+F \exp({G\omega})$. 
}\label{WC-relation}
\end{figure*}

\begin{figure*}[h]
\begin{center}
\includegraphics[width=10cm]{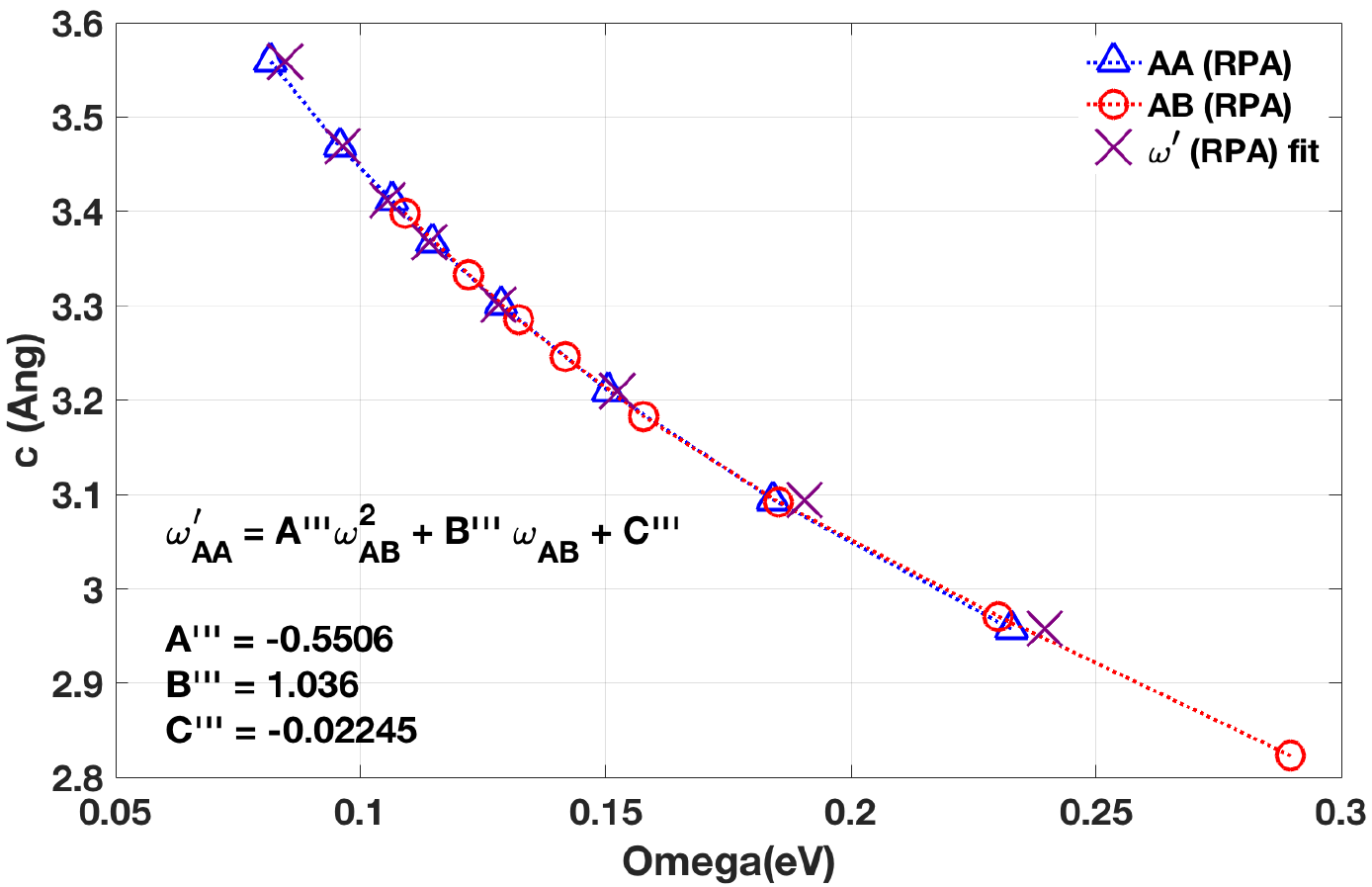}
\end{center}
\caption{
(Color online) Interlayer coupling strength ($\omega$, eV) versus interlayer distance (c, $\AA$) for different stacking geometries AA, AB (equivalent to BA) for equilibrium distances calculated within EXX+RPA calculations. Interlayer coupling strength ($\omega^{\prime}$, eV) of AA stacking is represented in-terms of interlayer coupling strength ($\omega$, eV) of AB stacking with a simple quadratic equation, $\omega^{\prime}_{AA} = A^{\prime\prime\prime}\omega^2_{AB}+B^{\prime\prime\prime}\omega_{AB}+C^{\prime\prime\prime}$. 
}\label{WC-relation2}
\end{figure*}

\begin{figure*}[h]
\begin{center}
\includegraphics[width=10cm]{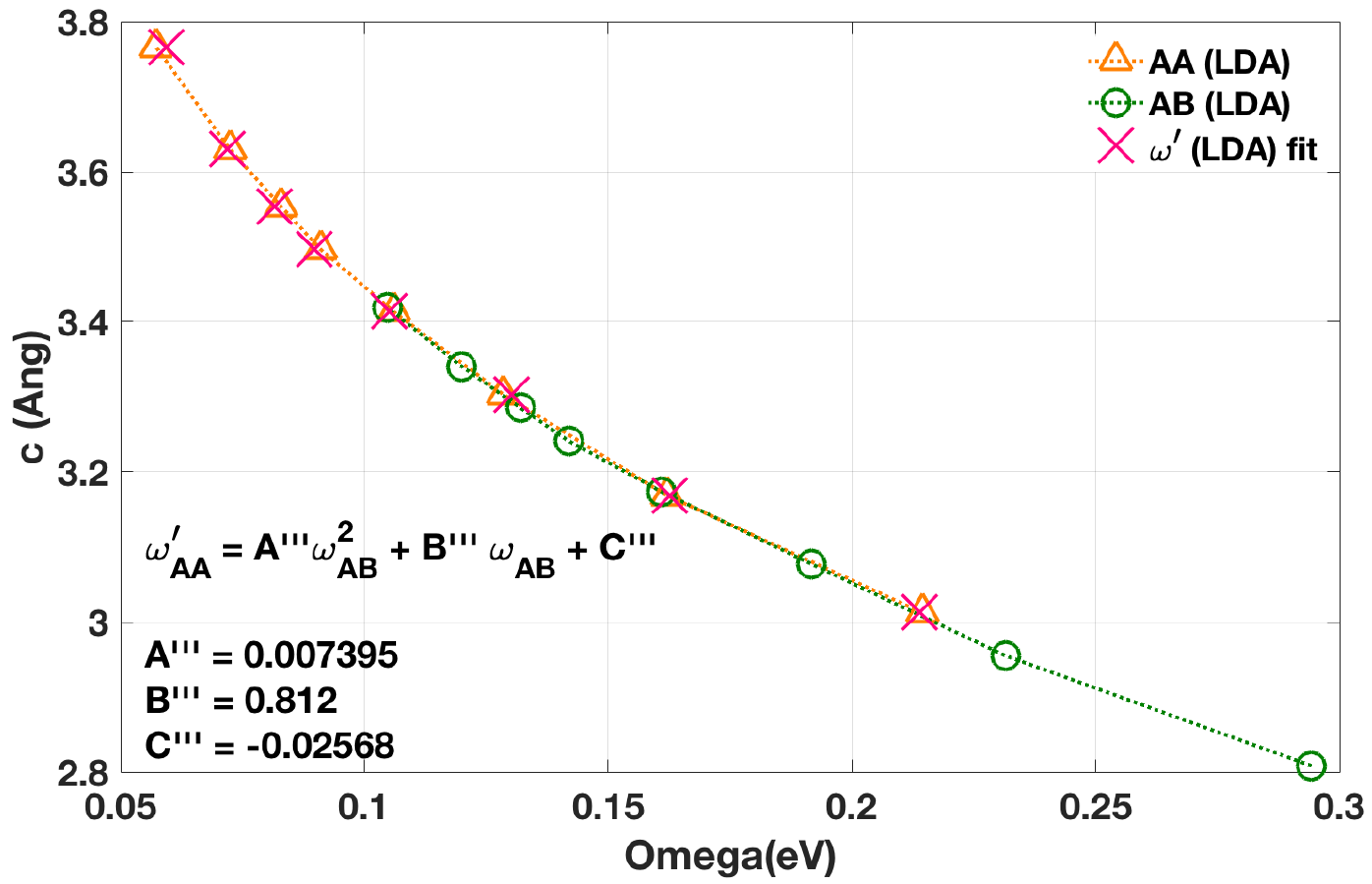}
\end{center}
\caption{
(Color online) Interlayer coupling strength ($\omega$, eV) versus interlayer distance (c, $\AA$) for different stacking geometries AA, AB (equivalent to BA) for equilibrium distances calculated within LDA calculations. Interlayer coupling strength ($\omega^{\prime}$, eV) of AA stacking is represented in-terms of interlayer coupling strength ($\omega$, eV) of AB stacking with a simple quadratic equation, $\omega^{\prime}_{AA} = A^{\prime\prime\prime}\omega^2_{AB}+B^{\prime\prime\prime}\omega_{AB}+C^{\prime\prime\prime}$. 
}\label{WC-relation3}
\end{figure*}

\clearpage

\subsection{Low energy effective model}    
The low energy effective model formulation results in simpler expressions for the Hamiltonian and often allows to 
obtain solutions with compact analytical forms. 
In the main text we have presented the best results for the full bands model but here we 
present for sake of completeness 
some analysis of the low energy bi-bilayer models subject to an electric field. 
The discrepancies between the results indicates that the low energy model is not useful to guide
experimental efforts with predictive accuracy. 
The low energy model hamiltonian for the twisted bi-bilayer with top ($+$) and bottom ($-$) bilayers is given by:
\begin{equation}  
H_{eff}(k)=
  \begin{pmatrix}
    \Delta_{1} & {h^+}(k)  & 0 & 0 \\
    {{h^+}{k}}^{\dag}  & -\Delta_{2}  & h_t & 0 \\
    0 & h_t^{\dag}  & \Delta_{2} &   {h^-}(k)\\
    0 & 0 &  {{h^-}(k)}^{\dag} &-\Delta_{1}\\
  \end{pmatrix},
  \label{hamiltonian}
\end{equation}
where
\begin{equation}
{h^{\pm}}(k)=\dfrac{\upsilon_{\rm F}^{2}}{3\omega}\left[(k_{x}-i\,k_{y})\,e^{i\,\theta/2} \right]^{2}
\end{equation}
and the site potentials for each layer are given by the parameter $\Delta_i$ where $i=1,\,2$ from top and bottom layers in each Bernal stacked bilayer. The tunneling matrix are as below:
\begin{equation}
H_{t} =   
\omega\begin{bmatrix}
     0 & 0  \\
     1 & 0  \\
  \end{bmatrix} , 
\omega\begin{bmatrix}
     0 & 0  \\
     e^{-i\,2\pi/3} & 0  \\
  \end{bmatrix},
\omega\begin{bmatrix}
     0 & 0  \\
     e^{i\,2\pi/3} & 0  \\
  \end{bmatrix}
\end{equation}
\begin{figure}
\begin{center}
\includegraphics[width=14cm]{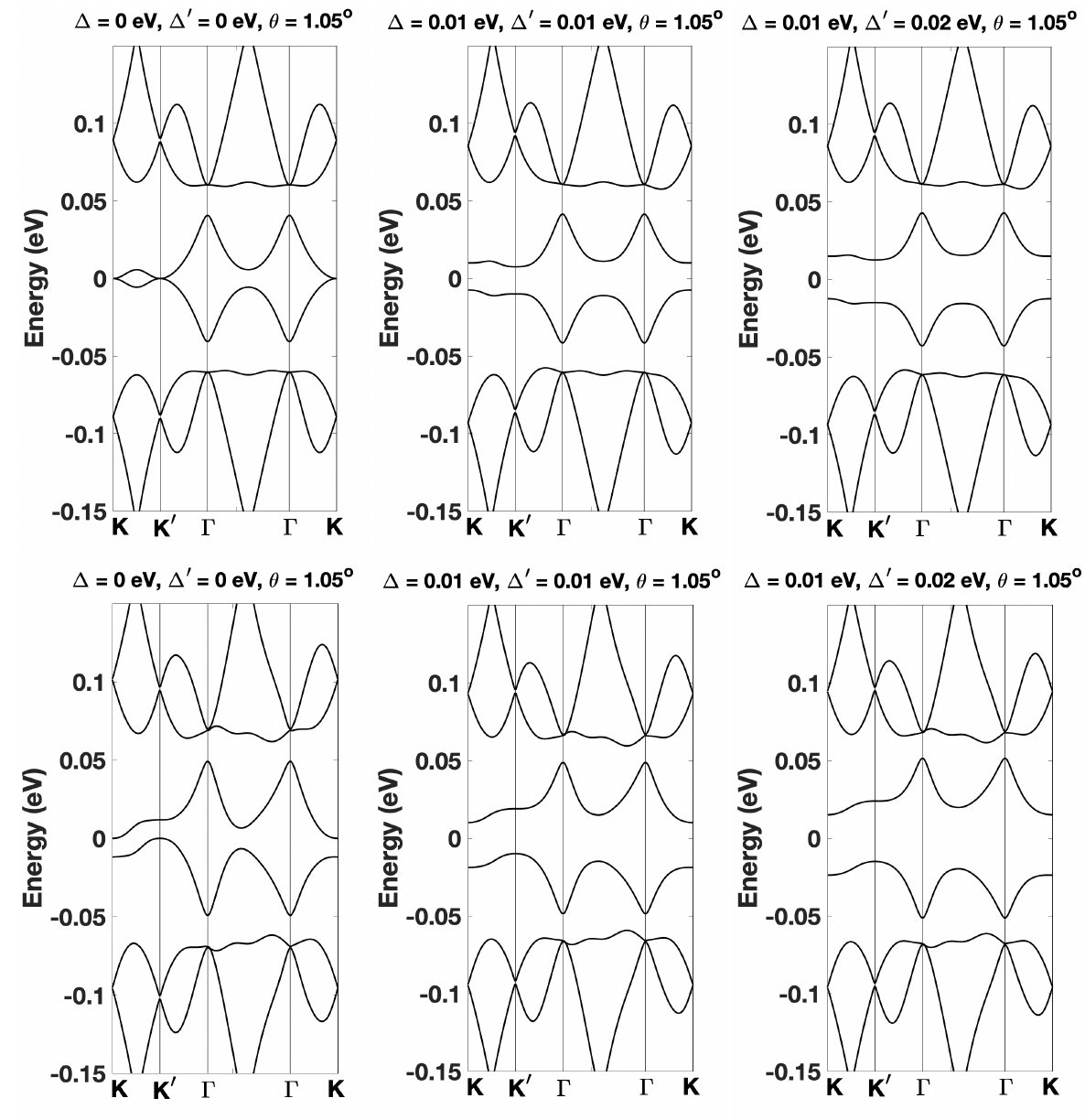}
\end{center}
\caption{
(Color online) The low energy effective model bandstructures obtained for the $\omega = 0.1 eV$ at twist angle $\theta = 1.05^o$ in three cases (left:) $\Delta^{\prime}=\Delta=0 $~eV, (middle:) $\Delta^{\prime}=\Delta=0.01 $~eV and (right:) $\Delta^{\prime}=2\Delta=0.02 $~eV. The bandwidth of effective low energy model is found to be larger compared to the eight band model carried out with Eq.~\ref{hamiltonian}
}
\label{lowBS}
\end{figure}
\begin{figure}
\begin{center}
\includegraphics[width=14cm]{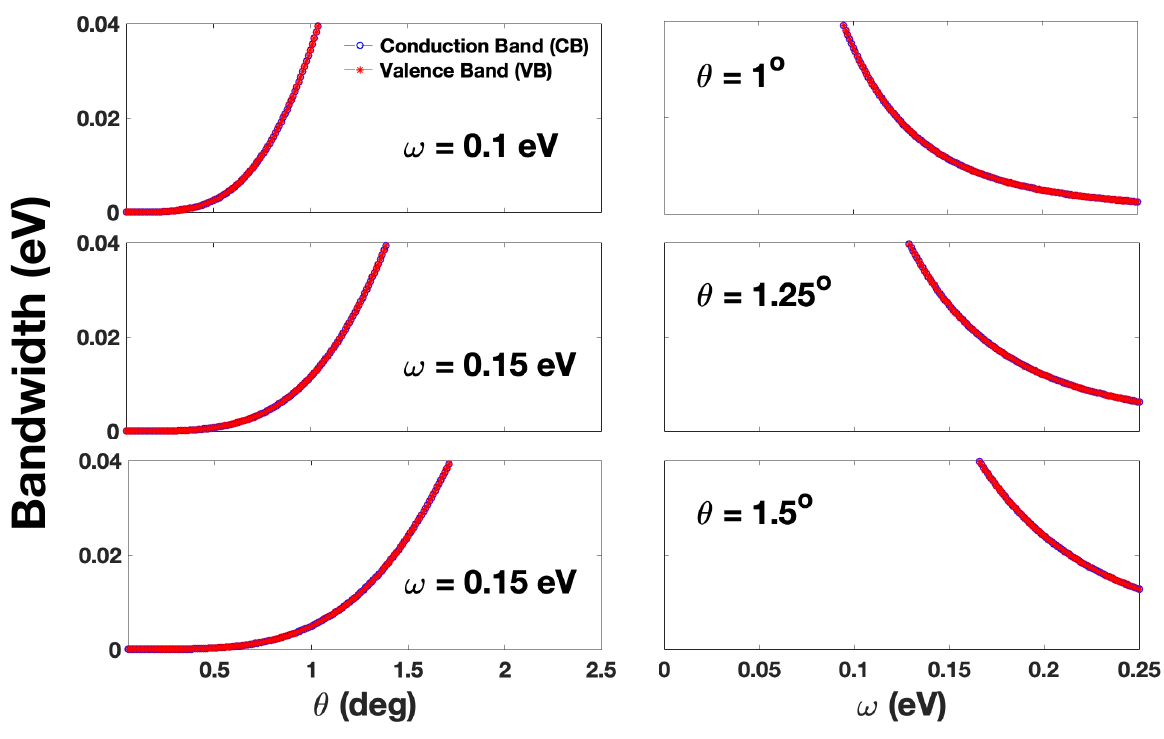}
\end{center}
\caption{
(Color online) The low energy effective model bandwidth variation for $\Delta^{\prime}=\Delta=0 $~eV as a function of inter-layer coupling strength and twist angle for the rigid continuum model with $t_0 = -2.6$~eV. {\em Left Panel:} Flatband bandwidth as a function of twist angle $\theta$ at different inter-layer couplings 
$\omega = 0.1, 0.15, 0.2$~eV. We observe that the bandwidth changes monotonically with the twist angle unlike the full bands model. 
{\em Right Panel:} Flatband bandwidth as a function of $\omega$, at different twist 
angles $\theta^{\circ}=1^{\circ}, 1.25^{\circ}, 1.5^{\circ}$.
We can observe a steep initial reduction in the bandwidth and no bumps for increasing $\omega$ like in the full bands model.
}
\label{lowBW}
\end{figure}

\subsection*{Expression for bandwidth}

Type-I twisted BGBG (AB stacked BG on AB stacked BG) Hamiltonian can be written as,
\begin{equation}
H = \sum_{a=t,\,b}\Psi_{a}^{\dagger}(\bold{k})H^{a}(\bold{k},\theta)\Psi_{a}(\bold{k}) \,+\, \sum_{j=0,\,\pm}\left( \Psi_{t}^{\dagger}(\bold{k})T_{j}\Psi_{b}(\bold{k}+\bold{Q_{j}})\,+\, h.c \right)
\end{equation}
where, 
\begin{equation} 
\begin{small} 
H^{t}(k,\theta)= 
  \begin{pmatrix}
    \Delta_{1} & \upsilon_{0}\,(\xi\,k_x-i\,k_y)e^{-i\theta/2}  & -\upsilon_{4}\,(\xi\,k_x-i\,k_y)e^{-i\theta/2} & \upsilon_{3}\,(\xi\,k_x+i\,k_y)e^{-i\theta/2} \\
    \upsilon_{0}\,(\xi\,k_x+i\,k_y)e^{i\theta/2}  &  \Delta_{1} & t_{\perp} & -\upsilon_{4}\,(\xi\,k_x-i\,k_y)e^{-i\theta/2} \\
    -\upsilon_{4}\,(\xi\,k_x+i\,k_y)e^{i\theta/2} & t_{\perp} &  -\Delta_{2} & \upsilon_{0}\,(\xi\,k_x-i\,k_y)e^{-i\theta/2} \\
    \upsilon_{3}\,(\xi\,k_x-i\,k_y)e^{i\theta/2} & -\upsilon_{4}\,(\xi\,k_x+i\,k_y)e^{i\theta/2} & \upsilon_{0}\,(\xi\,k_x+i\,k_y)e^{i\theta/2}  & -\Delta_{2}\\
  \end{pmatrix}\nonumber
  \end{small}
\end{equation}
\\
\begin{equation}  
\begin{small} 
H^{b}(k,\theta)=
  \begin{pmatrix}
    \Delta_{2} & \upsilon_{0}\,(\xi\,k_x-i\,k_y)e^{i\theta/2}    & -\upsilon_{4}\,(\xi\,k_x-i\,k_y)e^{i\theta/2} & \upsilon_{3}\,(\xi\,k_x+i\,k_y)e^{i\theta/2} \\
    \upsilon_{0}\,(\xi\,k_x+i\,k_y)e^{-i\theta/2}  & \Delta_{2} & t_{\perp} & -\upsilon_{4}\,(\xi\,k_x-i\,k_y)e^{i\theta/2} \\
    -\upsilon_{4}\,(\xi\,k_x+i\,k_y)e^{-i\theta/2} & t_{\perp}   &  -\Delta_{1} & \upsilon_{0}\,(\xi\,k_x-i\,k_y)e^{i\theta/2} \\
    \upsilon_{3}\,(\xi\,k_x-i\,k_y)e^{-i\theta/2} & -\upsilon_{4}\,(\xi\,k_x+i\,k_y)e^{-i\theta/2} & \upsilon_{0}\,(\xi\,k_x+i\,k_y)e^{-i\theta/2}  &  -\Delta_{1} \\
  \end{pmatrix} \nonumber
  \end{small} 
\end{equation}
\\
\begin{equation}  
T_{j}\simeq 
  \begin{pmatrix}
    0 & 0    & 0 & 0 \\
    0 & 0    & 0 & 0 \\
    0 & 0    & 0 & 0 \\
    3\omega & 0 & 0 & 0 \\
  \end{pmatrix}\nonumber
\end{equation}
\\
where, $\xi=\pm1$, for $K,\,K^{\prime}$ valleys, $\Psi_{b} = \{\phi_{B_{1}^{b}}\, , \phi_{A_{1}^{b}}\, , \phi_{B_{2}^{b}}\, , \phi_{A_{2}^{b}} \}^{\dagger}$, $\Psi_{t} = \{\phi_{A_{1}^{t}}\, , \phi_{B_{1}^{t}}\, , \phi_{A_{2}^{t}}\, , \phi_{B_{2}^{t}}\}^{\dagger}$ and $t_{\perp} =3\omega$ is the vertical hopping strength between the dimmer cites in bilayers and $T_{j}$ is the interlayer hopping matrix valid in the moderate twist angle limit. For analytical discussions, we have derived the Effective two band model Hamiltonian of tBGBG-I by integrating out the higher energy dimmer sites using the perturbation method,
\\
\begin{equation}  
H_{eff} =
  \begin{pmatrix}
    \Delta_{1} & \dfrac{-1}{(3\omega)^{3}}\left[(3\omega)^{2}\upsilon_{3}^{2}\,q^{2} \,+\, \upsilon_{F}^{4}q^{4}\,-\,6\omega \upsilon_{3}\upsilon_{F}^{4}q^{4}   \right] \\
    \dfrac{-1}{(3\omega)^{3}}\left[(3\omega)^{2}\upsilon_{3}^{2}\,{q^{\dagger}}^{2} \,+\, \upsilon_{F}^{4}{q^{\dagger}}^{4}\,-\,6\omega \upsilon_{3}\upsilon_{F}^{4}{q^{\dagger}}^{4}   \right]  & -\Delta_{1}  \\
  \end{pmatrix}
\end{equation}
\\
where, $q\,=\,\left[k^{2}-( \Delta K/2)^{2}  \right]\,; \,\, |\Delta K| = K_{\theta/2}-K_{-\theta/2}=\dfrac{4\pi}{3a}2\,sin(\theta/2)$ is the momentum shift between the Dirac cones of top and bottom bilayers due to rotation and $a=\sqrt{3}a_{cc}$. For simplicity we have taken $\upsilon_{4}=0$. The eigen energies of the above effective Hamiltonian are given by,
\\
\begin{equation}
\lambda^{\pm} = \pm \sqrt{ \Delta_{1}^{2} +\left|\dfrac{1}{(3\omega)^{3}} \left[ (3\omega)^{2}\upsilon_{3}^{2}q_{m}^{2}+\upsilon_{F}^{4}q_{m}^{4}-6\omega \upsilon_{3}\upsilon_{F}^{2}q_{m}^{4} \right]\right|^{2}  },
\end{equation}
\\
where, $q_{m}=|\Delta K|/2$ is the maximum value of $q$ at the Dirac points. The difference between the maximum and minimum values of the band energy is defined as the Bandwidth of the corresponding band. Here we have calculated the bandwidth,
\begin{equation}
W = (\lambda^{max}-\lambda^{min})\simeq \dfrac{1}{(3\omega)^{3}} \left[ (3\omega)^{2}\upsilon_{3}^{2}\dfrac{|\Delta K|^{2}}{4}+\upsilon_{F}^{4}\dfrac{|\Delta K|^{4}}{16}-6\omega \upsilon_{3}\upsilon_{F}^{2}\dfrac{|\Delta K|^{4}}{16} \right]-\Delta_{1} \nonumber
\end{equation}
From the above expression one can observe that the band width is decreasing nonlinearly with the interlayer coupling $\omega$ and external electric field is reducing the band width and trigonal warping strength is enhancing it within the limit of moderate electric field.  
\subsection{Magic twist angle}  
By equating the band width $(W)$ to zero we can get the magic angle, where it leads the formation of flatbands.
\\
\begin{equation}
(6\omega\, \upsilon_{3}\upsilon_{F}^{2}-\upsilon_{F}^{4}) |\Delta K|^{4} -4(3\omega)^{2} \upsilon_{3}^{2} |\Delta K|^{2} + 16 (3\omega)^{3} \Delta_{1} = 0
\end{equation} 

\end{widetext}


\begin{thebibliography}{59}
\bibitem{footnote}
In this work, we loosely use the term flatbands to refer to the sufficiently narrow bandwidth low energy bands where an effective Coulomb interaction $U$ versus bandwidth $W$ ratio satisfies $U/W \gtrsim 1$.
\end{thebibliography}
\end{document}